\pdfoutput=1
\documentclass[pageno,hyperref]{jpaper}
\usepackage[normalem]{ulem}
\usepackage{xspace} 
\usepackage[nocompress]{cite}
\usepackage{amsmath,amssymb,amsfonts}
\usepackage{braket}
\usepackage{tabularx,tabulary}
\usepackage{booktabs}
\usepackage{caption}
\usepackage{subfig}
\usepackage{comment}
\usepackage{balance}

\usepackage{color}
\usepackage[most]{tcolorbox}

\def\BibTeX{{\rm B\kern-.05em{\sc i\kern-.025em b}\kern-.08em
    T\kern-.1667em\lower.7ex\hbox{E}\kern-.125emX}}

\newcommand{\apps}{\textit{QUATRO}\xspace}

\begin{document}
\renewcommand{\hbar}{\mathchar'26\mkern-7mu h}
\renewcommand\footnotemark{}
\title{The \apps Application Suite: Quantum Computing for Models of Human Cognition}
\author{Raghavendra Pradyumna Pothukuchi\textsuperscript{1}, 
Leon Lufkin\textsuperscript{1}\textsuperscript{\textdagger}, 
Yu Jun Shen\textsuperscript{1}\textsuperscript{\textdagger}\textsuperscript{*}, 
Alejandro Simon\textsuperscript{1}\textsuperscript{\textdagger}\textsuperscript{*}, \\ 
Rome Thorstenson\textsuperscript{1}\textsuperscript{\textdagger}, 
Bernardo Eilert Trevisan\textsuperscript{1}\textsuperscript{\textdagger}\textsuperscript{*}, 
Michael Tu\textsuperscript{1}\textsuperscript{\textdagger}, 
Mudi Yang\textsuperscript{1}\textsuperscript{\textdagger}\textsuperscript{*},  
Ben Foxman\textsuperscript{1}\textsuperscript{\textdaggerdbl}, \\  
Viswanatha Srinivas Pothukuchi\textsuperscript{2}\textsuperscript{\textdaggerdbl},  
Gunnar Epping\textsuperscript{4}, 
Thi Ha Kyaw\textsuperscript{3}, 
Bryant J Jongkees\textsuperscript{5}, \\ 
Yongshan Ding\textsuperscript{1},
Jerome R Busemeyer\textsuperscript{4}, 
Jonathan D Cohen\textsuperscript{6}, 
Abhishek Bhattacharjee\textsuperscript{1}\thanks{\hspace{-6.5mm}\textsuperscript{\textdagger}Equal contribution. Listed in alphabetical order of last name. \\ \textsuperscript{\textdaggerdbl}Equal contribution. Listed in alphabetical order of last name. \\\textsuperscript{*}Worked as Yale University undergraduate.} }
\date{\begin{center}
    \textsuperscript{1}Yale University \textsuperscript{2}Independent  \textsuperscript{3}LG Electronics Toronto AI Lab \textsuperscript{4}Indiana University \textsuperscript{5}Leiden University \textsuperscript{6}Princeton University
\end{center}}

\maketitle

\begin{abstract}
    Research progress in quantum computing has, thus far, focused on a narrow set of application domains. Expanding the suite of quantum application domains is vital for the discovery of new software toolchains and architectural abstractions. In this work, we unlock a new class of applications ripe for quantum computing research---computational cognitive modeling. Cognitive models are critical to understanding and replicating human intelligence. Our work connects computational cognitive models to quantum computer architectures for the first time. We release \apps\footnote{\apps stands for quantum applications from cognitive modeling, and also means four, which is the number of cognitive model types we study.}, a collection of quantum computing applications from  cognitive models. The development and execution of \apps 
shed light on gaps in the quantum computing stack that need to be closed to ease programming and drive performance. Among several contributions, we propose and study ideas pertaining to quantum cloud scheduling (using data from gate- and annealing-based quantum computers), parallelization, and more. In the long run, we expect our research to lay the groundwork for more versatile quantum computer systems in the future.

\end{abstract}

\section{Introduction}
\label{intro}

Quantum computers are evolving from early experimental prototypes to more mature cloud platforms~\cite{awsQuantum,msQuantum,dwaveQuantum,ibmQuantum,Xanadu}, but their success now hinges on expanding their utility to more application domains. Unfortunately, most deployed quantum computing applications have thus far focused on select domains like physics, chemistry, finance, and recently, machine learning~\cite{quantumDiscovery}. This presents a chicken-and-egg problem. More general-purpose computing stacks would expand the benefits of quantum computing to more application domains. But, the paucity of new application domains has over-specialized quantum computing software to just a few application domains. New application domains are needed to shed light on more general-purpose quantum computing stacks of the future.

 In this work, we take a step towards breaking this vicious cycle of over-specialization by presenting computational cognitive modeling as a new target application domain for quantum computing. Cognitive models explain how humans process information and make decisions, and have been foundational to psychology, psychiatry, artificial intelligence (AI), machine learning (ML), and even economics~\cite{clsUpdate,pothos2022quantum,kahneman2003maps}.  These models are now adequately computationally demanding and attracting the attention of computer systems practitioners~\cite{distill}, making them a natural candidate for quantum computing research. Models used for relational reasoning~\cite{holyoak2012analogy} and planning~\cite{ho2020efficiency} rely on simultaneous constraint satisfaction kernels, which are known to be slow and scale poorly with model complexity. 
  
 Recently, cognitive scientists have also shown the prospect of using quantum probability theory to develop a unified and formal decision-making framework that explains human behavior difficult to be captured by existing classical alternatives, like biases, effects of question framing and order etc.~\cite{pothos2022quantum,rosendahl2020novel,rosendahlThesis}. Rigorous psychological studies with humans have shown evidence supporting the expressive power of quantum-theoretic frameworks and their improved modeling accuracy over existing methods~\cite{busemeyer2015bayesian,busemeyer2012quantum,pothos2013quantum,bruza2015quantum, pothos2022quantum,epping2023using,rosendahlThesis,rosendahl2020novel}. However, realizing the full potential of these methods requires developing and mapping larger models onto real quantum machines~\cite{pothos2022quantum}, a hitherto unexplored study.

Ours is the first work to map cognitive models to quantum hardware, offering a path to scale and augment them, and enable more human-like AI. We transcend prior work~\cite{ionqOrder}, which focuses only on small kernels or circuit components, to develop \apps, a suite of full, computationally hard classical and quantum-theoretic models, mapped to real quantum hardware. 

 Ours is also the first work to showcase algorithmic advances to realize constructs necessitated by cognitive models. Examples include constructs like subspace projections for quantum walks, implemented with new low-cost state-detection circuits. Although mapping cognitive models to quantum hardware drives the discovery of these constructs, they are broadly applicable including, e.g., quantum program assertions~\cite{assertNDD,assertSwap}.

Finally, by developing bridges between the cognitive sciences and quantum computing, we lay the foundations for new research in quantum computing software toolchains, programming languages, compiler support, control interfaces, reorganization of the quantum cloud, and more. These research questions go beyond the hardware-centric focus of existing work (e.g., error mitigation~\cite{Temme2017Nov,Endo2018Jul,Cai2022Oct}, reliability~\cite{veritas,qraft,mitigateBias,crosstalk,afs,q3de}, circuit synthesis~\cite{scaffcc,cutqc}, and microarchitecture~\cite{murali19insight}). To undergird our results, we evaluate new parallelism-aware cloud scheduling approaches on real quantum systems. Our approach offers performance speedups via parallelization and improved system throughput, representing new axes for performance scaling in quantum clouds. Like our newly discovered algorithmic constructs, these optimizations are broadly useful beyond cognitive modeling.

In summary, our specific research contributions are:
\begin{enumerate}
    \item We introduce \apps, end-to-end cognitive models as a new application domain for quantum computing. \apps is suitable for current and future quantum hardware.
    \item We discover new circuit-level constructs on subspace projections, techniques that benefit domains beyond cognitive modeling, with a fraction of resources over prior work. 
    \item We identify shortcomings and opportunities for research innovation in the programming stack for quantum computers. Unlike all prior work, ours transcends low-level hardware and addresses higher-level aspects of the quantum stack.
    \item We propose and analyze new quantum cloud organization using data from real quantum computers, showing 2.4$\times$--10$\times$ speedup for co-designed applications.
\end{enumerate}

Overall, our work expands quantum computing into a new high-impact application domain and sets the stage for building versatile quantum systems that are easier and faster to use. 

\section{Motivation}
\label{motiv}

Virtuous cycles among application domains, algorithms, and computer systems have been central to the decades-long success of computing. Consider, e.g., the advances in GPU and tensor hardware that have unlocked increasingly complex deep neural networks in the last decade~\cite{mittal2019survey}.
Unfortunately, quantum computing has yet to experience such rapid growth in application-systems co-design. Quantum applications are concentrated in physics~\cite{Georgescu2014Mar}, chemistry~\cite{Aspuru-Guzik2005Sep}, finance~\cite{Orus2019Nov,quantumFinance}, machine learning~\cite{quantumML}, and optimization~\cite{quantumOpt}. New applications are needed to ensure that the potential of quantum computing is more broadly harnessed by society~\cite{academiesQuantum,quantumDiscovery}.

We bridge the cognitive sciences with quantum computing as a step in enabling new families of quantum applications. Computational cognitive models describe how humans generate decisions and behavior~\cite{ratcliff1978,lca,musslickcohen2021}. Such insights have enabled breakthroughs in AI~\cite{clsUpdate}, ML~\cite{ntm,replayContinualLearn}, and behavioral economics that resulted in multiple Nobel prizes (Simon in 1978; Kahneman in 2002). 
Cognitive models are being studied today with the anticipation of identifying theoretical mechanisms to explain complex human behavior, and unlock AI that is not captured by modern deep learning~\cite{esbn,lake2017building}.

State-of-the-art cognitive models are computationally challenging, and are a natural target for quantum computing research. Models addressing fundamental human capabilities like relational reasoning and problem-solving involve simultaneous constraint satisfaction, and are realized with graph isomorphism techniques~\cite{holyoak2012analogy}, combinatorial optimization and planning~\cite{ho2020efficiency}, which are proven to be hard to scale classically. 

Furthermore, there is a new class of cognitive models that employs quantum probability as a theoretical tool to describe human decision-making. These models leverage the richer axioms of quantum probability such as projections, non-commutativity, and superposition, to describe complex behavior that has resisted explanation from all existing classical alternatives and/or identify parsimonious models with exponentially fewer parameters~\cite{pothos2022quantum,busemeyer2015bayesian}. Below are  two illustrative results.


As an example of the expressiveness of quantum cognitive models, consider a two-stage task where individuals choose to play or quit a gamble before each stage~\cite{tversky1992disjunction}. When the outcome of the first stage was known---regardless of whether it is a win or a loss, real-world data showed individuals generally preferred to play the gamble in the second stage. However, when the outcome was unknown, most individuals switched and prefer to quit, which violates the classic law of total probability, and so existing models cannot explain such ``irrational'' behavior.

Alternatively, a quantum cognitive model representing decision-making as a superposition over possible outcomes, predicts different results because of quantum interference effects that can generate classical violations of total probability~\cite{busemeyer2015bayesian}. 



Quantum cognition models also provide  lower dimensional representations of cognition and thus greater parsimony compared to classical models. Consider a  strategic game involving $n$ players, each choosing one of $K$ possible actions~\cite{bruza2015quantum}. Classically representing an individual's beliefs about the actions of all players requires $K^n$ joint probabilities, representing an exponential growth in model dimension. Instead, the same could be modeled with quantum probability using a single $K$-dimensional vector space, evaluated with $n$ different bases representing the  player's different view points of the game.

While there could be a future classical formalism that can match quantum cognition in expressiveness, model conciseness, and accuracy, quantum cognition provides some of the best empirically-validated tools today to explain complex behavior~\cite{rosendahl2020novel,busemeyer2012quantum,pothos2022quantum,rosendahlThesis,pothos2013can,pothos2013quantum,bruza2015quantum,kvam2015interference,epping2023using,busemeyer2006quantum,busemeyer2020comparison,busemeyer2015bayesian}. 
With commercial quantum computers being available, the timing is ideal to explore the use of quantum computers for large-scale modeling of human-level cognition.


Recently, a compilation framework, Distill~\cite{distill}, has been proposed to accelerate classical cognitive models on classical systems like multicore CPUs and GPUs. However, there is value in exploring how computationally hard cognitive models, including quantum cognitive models, can be mapped to quantum computers, which we discuss next.

\subsection{Why \apps Applications?}
\label{sec:why}

Our cognitive models are full-scale applications that introduce richer and newer constructs not found in architecture benchmarking kernels like SuperMarq~\cite{supermarq}. 
Such constructs include Hamiltonian-based walks, absorbing boundaries requiring projections, walks on analog quantum machines, multi-eigenstate problems, non-unitary eigensolvers, and nonlinear dynamics. These constructs are useful beyond cognitive models, e.g., in condensed matter physics~\cite{mallick2019simulating}, or dynamics of electron systems~\cite{oka2005breakdown}, but have received scant attention from architects.
 
 Historically, end-to-end applications enabled computer architects to discover innovation missed by simple benchmarks. For example, by offering computer architects insight into datacenter applications, CloudSuite~\cite{cloudsuite} unearthed shortcomings of CPU microarchitecture missed by simpler benchmarking suites like SPEC (e.g., inefficiencies in instruction fetch). Analogously, mapping the \apps applications and executing them, as we will present in Sections~\ref{models} and~\ref{results},  exposed many limitations of the quantum computing stack today inspiring new research. We had to innovate new state-detection circuits, nonlinear dynamics solvers, and multi-stage translators to map these applications, and discovered architecture-algorithm co-design opportunities like data-parallel RBM training, and parallel eigenstate computation to tap new axes of speedup. These are broadly applicable and advance quantum architecture.


Finally, \apps presents current cognitive models, factored for implementation on quantum platforms; enabling these to run on quantum computers promises substantial progress in scaling them to address more complex tasks than currently possible.  It could help identify how a quantum formulation could provide a more accurate and/or coherent account of human cognition, such as the stochasticity of decision-making, and the effects of cognitive arousal, than has been achieved classically~\cite{bruza2015quantum,rosendahlThesis}. It could also help answer how the brain solves seemingly computationally hard tasks, and whether mental events obey the same probability rules as physical events~\cite{bruza2015quantum}.

\begin{table*}[ht]
    \caption{The \apps applications: Selecting cognitive models for quantum implementation.} 
    \label{tab:models}
    \renewcommand*{\arraystretch}{1}
    \scriptsize
    \centering
    \begin{tabulary}{0.98\textwidth}{p{0.135\textwidth}p{0.05\textwidth}p{0.1\textwidth}p{0.18\textwidth}p{0.17\textwidth}p{0.2\textwidth}}
    \toprule
        \textbf{Model} & \textbf{Type} & \textbf{Purpose} & \textbf{Significance} & \textbf{Quantum methods we apply} & \textbf{Relevance for systems} \\ \toprule
        Quantum Walk~\cite{ratcliff2016diffusion,busemeyer2019cognitive} & Quantum-theoretic & Decision-making (2-choice) & Accuracy, size, and flexibility advantage over classical walks & Quantum walk simulation (gate-based, annealing) & Presents Hamiltonian based walks, and with absorbing boundaries\\ \midrule
        Multi-Particle Multi-Well (MPMW)~\cite{rosendahl2020novel,rosendahlThesis} & Quantum-theoretic & Decision-making (multi-choice) & Accuracy, size and flexibility advantage. First unified model over many cognitive parameters & Eigensolution: SSVQE, and QITE (both gate-based) & Studies multi-eigenstate problem, non-unitary algorithms \\ \midrule
        Predator-Prey~\cite{predprey} & Classical, hard & Cognitive control & Captures attention, cognitive resource utilization and planning & Restricted Boltzmann Machine (RBM; annealing) & Introduces computationally hard simultaneous constraint satisfaction, and brain-inspired neural networks \\ \midrule
        Leaky Competing Accumulator (LCA)~\cite{lca,lcaExtend} & Classical, hard& Decision-making and control & Biologically plausible model & Nonlinear dynamics (annealing) & Introduces simultaneous constraint satisfaction with nonlinear dynamics\\
        \bottomrule
    \end{tabulary}
\end{table*}
\begin{figure*}[t]
\centering
\subfloat[Quantum walk on a 1-dimensional lattice for two alternatives. 
The walk is described by Hamiltonian, $H$.]{
\includegraphics[width=0.27\textwidth]{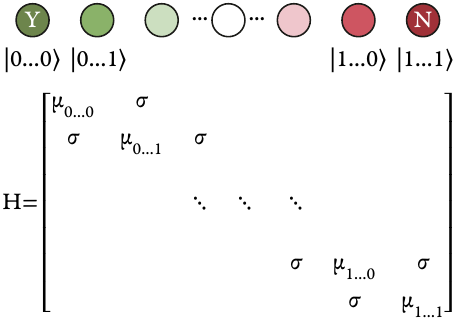}
\label{subfig_walks}
} \hspace{1mm}
\subfloat[MPMW for one infinitely tall, finite width well. Its lowest three eigenvectors are shown.]{
\includegraphics[width=0.24\textwidth]{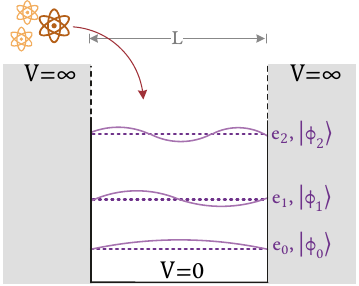}
\label{subfig_wells}
} \hspace{1mm}
\subfloat[Predator-Prey screen grid to model cognitive control/attention.]{
\includegraphics[width=0.18\textwidth]{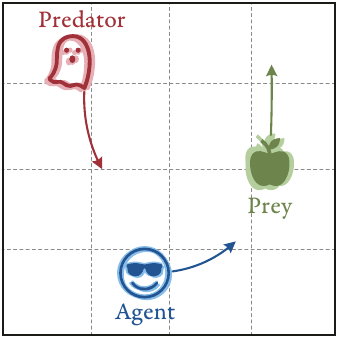}
\label{subfig_pp}
} \hspace{1mm}
\subfloat[LCA for two inputs. f$_1(.)$, f$_2(.)$ are nonlinear functions.]{
\includegraphics[width=0.19\textwidth]{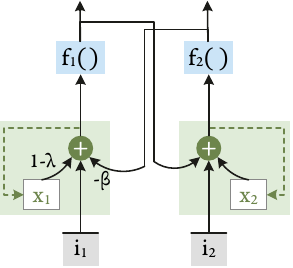}
\label{subfig_lca}
}
\caption{Cognitive model prototypes for which we develop quantum implementations.}
\label{fig_models}
\end{figure*}
\section{Relevant Quantum Techniques}
\label{algo}

In this section, we briefly review important quantum computing concepts relevant to the development of \apps. More details can be be found in standard texts (e.g.,~\cite{ding2020circuit,quantumCountry}).

We use the Dirac (bra-ket) notation~\cite{quantumCountry} ($\hbar=1$), where $\ket{x}$ denotes a column vector $x$  used to represent states, $\bra{x}$ is a row vector, and block letters (e.g., $H$) are operators. A quantum system is described by its Hamiltonian, $H$. If the system's initial state is $\ket{\psi_{init}}$, then its state after time $t$ is, $\ket{\psi_{new}} = U\ket{\psi_{init}}$, where $U=e^{-iHt}$ is the evolution operator. 

\vspace{1mm}
\noindent
\textbf{Quantum walks: } Quantum walks, which describe the evolution of a particle over some states \cite{farhi1998quantum,quantumWalkReview}, are used in cognitive models for decision-making~\cite{busemeyer2006quantum, kvam2015interference}. They capture how individuals process inputs (stimuli) to navigate their internal preferences (states) under uncertainty to make decisions. 



Quantum walks are typically implemented on gate-based computers (also called digital quantum computers) by compiling $U$ into gates~\cite{scaffcc,qiskit} and initializing the system to $\ket{\psi_{init}}$. The term $U$ is usually directly specified. Many specifications are possible, with coin-walks on graphs~\cite{quantumWalkReview} being a common approach. Cognitive models require, however, a different approach, with walks specified by $H$~\cite{busemeyer2006quantum, kvam2015interference}, and their execution on gate-based systems requires exponentiation to derive $U$. 


\vspace{1mm}
\noindent
\textbf{Variational algorithms: } Using classical and quantum computation for optimization, variational algorithms are important near-term constructs to find the eigenstates of cognitive models.

Variational quantum eigensolvers (VQEs)~\cite{Peruzzo2014Jul,variationalAlgo} and the Quantum Approximate Optimziation Algorithm (QAOA)~\cite{qaoa} are two important variational algorithms that minimize cost functions, thereby finding the lowest eigenstate (or ground state). Cost functions can be the expectation of the Hamiltonian or custom-defined.

SSVQE (Subspace VQE) is a recently proposed technique to identify eigenstates higher than the ground state~\cite{ssvqe}. These higher states are required by quantum cognitive models~\cite{rosendahl2020novel,rosendahlThesis}, unlike most other applications that only need the ground state.

Similar to VQE, SSVQE uses a parameterized quantum circuit (or ansatz) to generate candidate solutions for the optimization. Then, a classical optimizer evaluates the cost function for these states, and changes the circuit parameters to explore better states. However, SSVQE differs from VQE in computing the cost function since it must find excited states. 

We use two relevant SSVQE variants~\cite{ssvqe}. One approach (SSVQE B) finds one excited state at a time, e.g., the k$^{\mathrm{th}}$ state using a weighted cost function. At each iteration, the algorithm inputs $k+1$ orthogonal vectors, $\{\ket{\psi_j}\}_{j=0}^k$, into the ansatz (e.g., $k+1$ vectors whose elements are all zeros except for a 1 in a different position each), and obtains states $\{\ket{\phi_j}\}_{j=0}^{k}$, and their energy expectations $\bra{\phi_j}H\ket{\phi_j}$. Then, a classical optimizer minimizes the cost function with weight $w$,  $w\bra{\phi_k}H\ket{\phi_k} + \sum_{j=0}^{k-1}\bra{\phi_j}H\ket{\phi_j}$, to obtain the k$^{\mathrm{th}}$ eigenstate.


The other method (SSVQE C) also uses a weighted cost function but finds all eigenstates up to the $k^{\mathrm{th}}$ state simultaneously. The cost function is $\sum_{j=0}^{k}w_j\bra{\phi_j}H\ket{\phi_j}$, where the weights $w_j$ decrease in value i.e., $w_j < w_{j-1}$. 
As we will show, the two methods have different implementation tradeoffs.




\vspace{1mm}
\noindent
\textbf{Quantum imaginary time evolution (QITE): } Finding eigenvalues of cognitive models with variational algorithms requires identifying a good ansatz and optimizer configuration. Unfortunately, this is hard~\cite{ansatzChoice}. Therefore, we also consider an alternative approach, QITE, to obtain eigenvalues~\cite{qite}.

QITE works by evolving a quantum state with the time parameter as $t=-i\tau$. Then, the state is given by $e^{-H\tau}\ket{\psi_{init}}$, which represents state decay. If $\tau$ is long, all higher energy components vanish, and the system decays into the groundstate. To obtain excited states, we use the Quantum Lanczos method that uses several QITE states with different $\tau$ values~\cite{qite}. 

QITE belongs to the class of non-unitary algorithms that can provide quantum speedup, but require additional steps to run on quantum computers because the hardware is unitary~\cite{McArdle2019Sep,Kyaw2022Aug,Saxena2023Jun}. Such methods have not been studied in prior systems research. 



\vspace{1mm}
\noindent
\textbf{Quantum annealing: }
Cognitive models using simultaneous constraint satisfaction~\cite{quantumRBM,lca} and quantum walks can be cast as optimization problems solved by quantum annealing. In quantum annealing~\cite{quantumAnneal}, the system begins in the ground state of a simple Hamiltonian, which is slowly changed to the Hamiltonian whose minimum is to be found. If the rate of change is slow enough, the system ends up in the ground state of the target Hamiltonian, giving the solution we seek~\cite{annealOrig,Kyaw2014Oct,Kyaw2018Apr,quantumAnneal,Kyaw2017May,vanDam2001Oct}. 

Quantum annealers represent analog quantum computers, which are programmed directly with the Hamiltonian~\cite{dwaveQuantum,queraQuantum,simuq,quantumAnneal} unlike digital versions. This makes them attractive for cognitive models, which are commonly specified with the Hamiltonian, but such systems haven't received adequate attention in quantum computer architecture research.

Present quantum annealers take as input Ising Hamiltonians, which are of the form $H = \sum_{\substack{i,j}}a_{ij}{\sigma_z}^i\sigma_z^j + \sum_ib_i\sigma_z^i$, where $\sigma_z^i$ is the Pauli $Z$ spin operator~\cite{quantumAnneal} whose values can be +1 or -1, and $a_{ij}$,\,$b_i$ are scalars and $i$,\,$j$ refer to the qubits. General Hamiltonians, including those in cognitive models, additionally contain Pauli $X$ ($\sigma_x$), Pauli $Y$ ($\sigma_y$) operators. Annealing such Hamiltonians requires manual reformulation to map the applications to the hardware~\cite{pauliToIsing}.

\section{\apps: Quantum Cognitive Applications}
\label{models}

We identify important cognitive models that can benefit from quantum execution, and which also introduce new constructs that have not been studied in quantum systems research. Table~\ref{tab:models} lists these models, the probability theory they use, significance to cognitive science, the techniques we use to map them to quantum hardware, and the new features they present for systems research. We present our implementations next.





\subsection{Quantum Walk}
\label{sub:qwalk}


\textit{Quantum Walk} captures 2-choice decision-making (e.g., yes/no tasks) using the biased random walk of a particle on a 1-dimensional lattice of states~\cite{ratcliff2016diffusion,busemeyer2019cognitive}, as shown in Figure~\ref{subfig_walks}. The states represent an individual's preference levels for the choices, and the model is given by the Hamiltonian $H$. This model is flexible to describe many aspects of decision-making such as uncertainty, perspective; has better accuracy; and is more compact than classical walks~\cite{busemeyer2015bayesian,kvam2015interference}.


There are two variants of \textit{Quantum Walk}~\cite{busemeyer2006quantum}, one where the walk evolves over all states including the boundary states for a fixed number of steps (called reflecting boundaries), and another where the walk evolves only until it doesn't reach a boundary (called absorbing boundaries). The former represents decision-making responses to a prompt, and the latter captures the first time an individual arrives at a decision. 

\begin{figure}[h]
\vspace{-3mm}
\centering
\subfloat[Reflecting. ]{
\raisebox{4.5mm}{\includegraphics[width=0.35\linewidth]{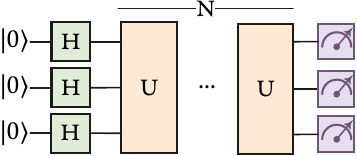}}
\label{subfig_rwalk}
} 
\subfloat[Absorbing. ]{
\includegraphics[width=0.58\linewidth]{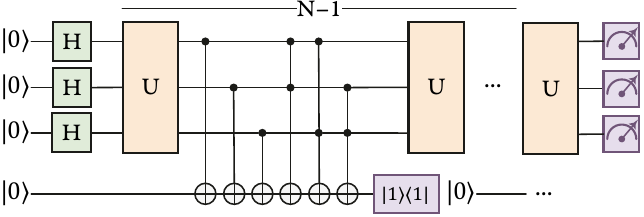}
\label{subfig_awalk}
} 
\caption{Realizing an 8-state Quantum Walk on gate-based systems. \textbf{H} is the Hadamard gate.}
\label{fig_walks}
\vspace{-4mm}
\end{figure}

\vspace{1mm}
\noindent
\textbf{Reflecting boundaries (gate-based): } 
The probability of a decision state (e.g., $ \ket{d} = \ket{0\ldots0}$) after $N$ timesteps is given by $||MU^N\ket{\psi_{init}}||^2_2$, where $M = \ket{d}\bra{d}$ is the measurement operator~\cite{busemeyer2006quantum}. Realizing this walk on a digital quantum computer is straightforward, using the circuit construction in Figure~\ref{subfig_rwalk} for an example 3-qubit (8-state lattice) model.

\vspace{1mm}
\noindent
\textbf{Absorbing boundaries (gate-based): } This walk evolves only in the non-boundary states, and requires projection after each timestep to contain its evolution. Thus, the probability for a decision after $N$ timesteps is given by $||MU(PU)^{N-1}\ket{\psi_{init}}||^2_2$, where $P$ is the projector for the non-boundary states.


Realizing this walk on existing hardware is difficult because the hardware doesn't support partial projections. The only approach is post-selection, i.e., running the system many times and discarding the trials where the system was found in the boundaries. This requires a non-disruptive, runtime mechanism to identify when the system touches the boundaries. 

Unfortunately, prior state-detection/checking circuits~\cite{proq,assertNDD,assertSwap} are ill-suited for our purpose. We cannot use projection-based methods~\cite{proq} since they change the state. Among alternatives, one approach~\cite{assertNDD} detects only certain states, such as those with an even number of ones. In our case, a boundary state can have any number of ones. The other, swap-based assertions~\cite{assertSwap}, can check for approximate state membership but doubles the number of qubits to swap and restore the state. 

To overcome these problems, we developed a new circuit to detect the boundary states with only one ancilla qubit. Figure~\ref{subfig_awalk} shows the circuit with our proposed state detector for an example 8-state walk. The ancilla qubit is shown at the bottom. For any input state $\sum_{i=0}^7a_i\ket{i}\otimes \ket{0}$, the detector's output is $(a_0\ket{0}+a_7\ket{7})\otimes\ket{0}+(\sum_{i=1}^6a_i\ket{i})\otimes \ket{1}$ i.e., the ancilla is in $\ket{1}$ if and only if the main qubits are not in the boundary states. Therefore, post-selecting on this condition is sufficient to measure the correct probabilities. The detector is easily extended to larger systems and other states by varying the CNOT and Toffoli gates. This is the first generic absorbing boundary walk design on existing quantum computers.

\vspace{1mm}
\noindent
\textbf{Annealing walk: } Digital quantum computers are programmed with $U$, which requires exponentiating the $H$ given by the cognitive model. To avoid this expensive step, we study \textit{Quantum Walk} on analog computers like quantum annealers~\cite{dwaveQuantum}.  Unfortunately, existing annealers accept only Ising Hamiltonians~\cite{dwaveQuantum}, which have diagonal elements and can be decomposed into pairwise Pauli-Z terms ($\sigma_z$). The \textit{Quantum Walk} Hamiltonian is not diagonal, resulting in additional Pauli-X ($\sigma_x$) and Pauli-Y ($\sigma_y$) terms, and can have more than two Pauli operator interactions. Therefore, we need a new approach.

\begin{figure}[h]
\centering
\includegraphics[width=0.8\linewidth]{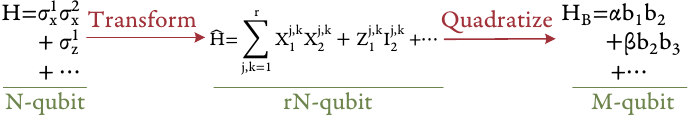}
\caption{Realizing \textit{Quantum Walk} on annealing systems.}
\label{fig_annealwalks}
\vspace{-1mm}

\end{figure}

We build on prior work~\cite{pauliToIsing} that mapped molecular Hamiltonians to annealers, to map our \textit{Quantum Walk} model. The multi-stage translation we use is shown in Figure~\ref{fig_annealwalks}. We begin by decomposing the N-qubit \textit{Quantum Walk} Hamiltonian into a sum of Pauli interactions. Next, we create a larger rN-qubit system and map these interactions to the larger system using new operators (e.g., X) that only use Pauli-Z operations. 
Finally, we convert the Pauli-Z operations in the new system to binary variables, and quadratize them i.e., expand the Hamiltonian such that each term has at most two variables~\cite{quadratization}. The result, $H_B$, can be annealed with existing machines, and the original ground state can be recovered from the annealer's output. Increasing the value of $r$ results in more accurate estimates.


Even though the annealing version of \textit{Quantum Walk} only gives the ground state instead of full dynamics, it is still useful for cognitive scientists. Mapping this model also highlights the challenges in using Hamiltonian-based systems today.

\subsection{Multi-Particle Multi-Well (MPMW)}
\label{sub:mpmw}

The \textit{MPMW} is a quantum cognitive model for multi-choice decision-making~\cite{rosendahl2020novel,rosendahlThesis}. It uses a 1-dimensional landscape of potential wells, one for each choice. The well's parameters like its height, width, and separation from neighbors, correspond to the cognitive parameters of attention, internal representation and concept similarity, respectively. Figure~\ref{subfig_wells} shows the simplest \textit{MPMW} model with one infinite height well~\cite{rosendahlThesis}. The model uses particles with different energies, representing an individual's cognitive arousal, to describe how individuals integrate information. Each particle enters the landscape serially and its final position is measured. When a particle falls within a well, the information towards that choice is incremented by one bit. This process repeats until a definitive choice is made.

The MPMW is the first formal, unified model to integrate several cognitive parameters, and outperforms classical alternatives in accuracy, model conciseness, and expressiveness~\cite{rosendahlThesis}. 


In \textit{MPMW}, the particle's position is determined by the eigenstates of the landscape's Hamiltonian, and finding them is the key computational step. We begin by deriving the Hamiltonian from the Schr\"{o}dinger equation of the particle in the landscape. In this equation, we discretize the position dimension, and use the finite difference method for the derivative to obtain the Hamiltonian operators, whose eigenstates we then find.


We identify two approaches for eigensolving: SSVQE~\cite{ssvqe}, which is variational, and QITE/Quantum Lanczos~\cite{qite}, which is not. We select these methods because they demand different design and execution support on quantum hardware. 

\vspace{1mm}
\noindent
\textbf{SSVQE: }We select two variants, SSVQE B and SSVQE C (Section~\ref{algo}) that present different algorithmic-implementation tradeoffs. SSVQE B runs a simpler optimization to find one eigenstate, but must be run multiple times to find multiple eigenstates. Instead, SSVQE C runs a single but complex optimization to find all eigenstates. While SSVQE C is more algorithmically efficient, SSVQE B is embarrassingly parallel and each instance is faster than SSVQE C. This presents a new opportunity for speedup that could be tapped with an efficient system architecture. This axis of parallelism is also complementary to existing methods that parallelize the expectation calculation of Pauli operators within a variational step~\cite{simultaneousPauli,simultaneousPauli2}, or those that parallelize runs for averaging~\cite{mineh2023accelerating}.


\vspace{1mm}
\noindent
\textbf{QITE/Quantum Lanczos: } Variational algorithms like SSVQE require selecting a suitable ansatz and optimizer, which is generally difficult~\cite{ansatzChoice}. 
Hence, we pick an alternative non-variational algorithm, QITE~\cite{qite}, for eigensolving. QITE is in the class of non-unitary quantum algorithms \cite{McClean2017Apr,Kyaw2022Aug,Saxena2023Jun}, which haven't been studied in systems design. Being non-unitary, QITE is unsupported natively on quantum computers, which are unitary. Therefore, we follow prior work~\cite{qite} to determine unitary operators that result in the same state evolution as with QITE. At each timestep, these operators are obtained by solving a linear system of equations with Pauli-operator expectations, which in turn, are measured from a quantum computer.


To find the excited states, we use the Quantum Lanczos algorithm base on QITE~\cite{qite}.  This approach obtains different state vectors evolved with QITE to construct a vector subspace. Then, it runs a classical Lanczos iteration on this subspace to orthogonalize the states into excited eigenstates.

\subsection{Predator-Prey}
\label{sub:pred-prey}

\textit{Predator-Prey} is used to model cognitive control~\cite{predprey}, which 
governs other mental processes including decision-making. The model describes a player  playing a game shown in Figure~\ref{subfig_pp}. The game has a screen grid with three entities: an agent, which is the player's screen icon, a prey, and a predator. The player must move the agent to capture the prey and avoid being caught by the predator. The model captures how the player allocates attention to the screen icons to discern their position and uses those perceived positions to make a move.

\begin{figure}[h]
\centering
\includegraphics[width=0.6\linewidth]{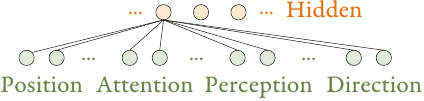}
\caption{Quantum RBM network for Predator-Prey.}
\label{fig_qRBM}
\vspace{-3mm}
\end{figure}

Identifying the optimal attention levels allocable to each icon, and deciding movement is a simultaneous constraint satisfaction problem.
We model it with a quantum restricted Boltzmann machine (RBM) that we propose~\cite{quantumRBM}, shown in Figure~\ref{fig_qRBM}. We choose an RBM because it aligns with the simultaneous interactions found in the brain, and also for its deep connection with cognitive modeling~\cite{smolensky1986information,hinton2007learning}. Our RBM has a visible layer with nodes for the true position of the on-screen entities, attention, perceived positions, and the movement direction.  

We map our RBM to a quantum annealer since it is a good fit for the RBM's Ising interactions. The standard approach to run the RBM uses 2 anneals for inference (visible to hidden, and back) and 3 for training (+1 to update hidden nodes). 
We additionally propose two new methods that benefit from different system architectures. One approach (RBM$_\textrm{eff}$) leverages compute-parallelism by combining updates to both hidden and visible layers. This method is resource efficient, requiring only 1 anneal for inference and 2 for training, but has more complexity. Another approach  (RBM$_\textrm{parK}$) leverages data-parallelism where we anneal for $K$ different samples simultaneously. This is inspired by asynchronous ML training algorithms~\cite{hogwild}, where the network weights are updated once after $K$ samples, instead of $K$ sequential updates.

\subsection{Leaky Competing Accumulator (LCA)}
\label{sub:lca}

The \textit{LCA} is a biologically inspired model for multi-choice decision-making and cognitive control~\cite{lca,lcaExtend}. Figure~\ref{subfig_lca} shows the \textit{LCA} for two choices. Each choice has an accumulator $x(t)$, which integrates the values of the input $i(t)$ corresponding to that choice at timestep $t$. The accumulator loses or leaks its value (by a factor $\lambda$), and is inhibited by the output of the other choice (by a factor $\beta$), i.e., $x_1(t) = i_1(t) + (1-\lambda)\cdot x_1(t-1) -\beta f_2(t)$, where $f()$ is a nonlinear function such as a sigmoid that acts on the accumulator's output.

 We select the \textit{LCA} because it is widely used, and its mathematical behavior has been well-characterized~\cite{lcaExtend}. Moreover, the \textit{LCA} is a simultaneous constraint satisfaction model like Ising models, with the additional complexity of stateful dynamics or memory---occurring due to the accumulator, and nonlinearity---due to the activation function. These aspects have been crucial in its utility as a model, but they also create challenges in running it on existing machines.

We map the \textit{LCA} to quantum annealers since they can solve Ising problems. We first linearize $f()$ into the format $ax+b$ using its Taylor expansion. Next, to obtain the value of $f()$ for a given value of $x$, we use the annealing cost function, $(f_x -(ax+b))^2$, following prior work that used annealing for prime factorization~\cite{pauliReduce}. 
Minimizing this cost function would yield $f_x$ to be $\approx f(x)$. Then, we replace $x$ with its previous value ($x(t-1)$), and the input. This gives a cost function with  $f_x$, $x(t-1)$, and the inputs $i(t)$. Next, we expand each variable with multiple qubits to represent floating point numbers. Finally, we quadratize the cost function before annealing. 

We explored different alternatives to run the \textit{LCA} over multiple timesteps. One approach is to anneal for a single timestep at once, where the outputs from the previous timestep are used as inputs for the next anneal. Since existing annealers do not offer easy variable initialization, we use additional terms in our cost function to realize it. This method has the overhead of reading and re-initializing qubits at each step.

Another approach is to unroll the \textit{LCA} dynamics for $K$ steps, so that one anneal would return the output for all $K$ steps. This amortizes the reading/initialization overheads but results in a much larger Hamiltonian than what is reliably run on current systems.  The last approach uses the technique of Feynman's clock, where a new problem is formulated that includes a timestep register along with the \textit{LCA} variables~\cite{feynmanClock}.  This method introduces complex variables that are unsupported presently. As a balance, we anneal for $K$ steps at once. 

\section{Experimental Setup}
\label{setup}


\begin{figure*}[t]
\centering
\subfloat[Timestep 0.]{
\includegraphics[width=0.19\textwidth]{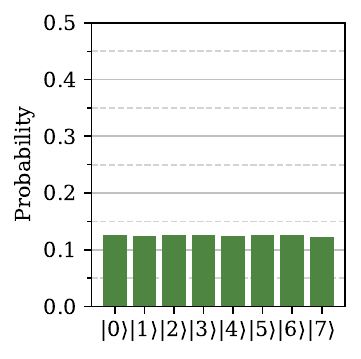}
\label{subfig_rwalks_qasm0}
} 
\subfloat[Timestep 1.]{
\includegraphics[width=0.18\textwidth]{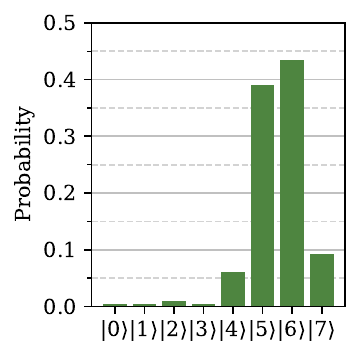}
\label{subfig_rwalks_qasm1}
} 
\subfloat[Timestep 2.]{
\includegraphics[width=0.18\textwidth]{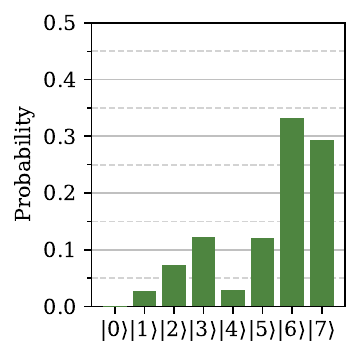}
\label{subfig_rwalks_qasm2}
} 
\subfloat[Timestep 3.]{
\includegraphics[width=0.18\textwidth]{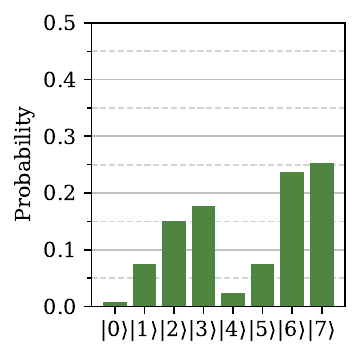}
\label{subfig_rwalks_qasm3}
}
\subfloat[Timestep 4.]{
\includegraphics[width=0.18\textwidth]{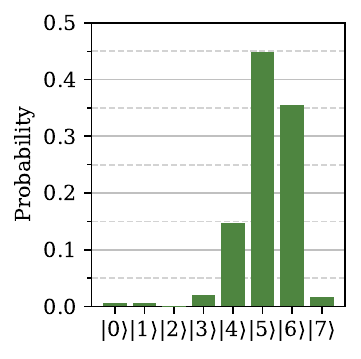}
\label{subfig_rwalks_qasm4}
}
\caption{Reflecting boundary \textit{Quantum Walk} with statevector simulation.}
\label{fig_refwalk_qasm}
\vspace{-5mm}
\end{figure*}

\begin{figure*}[t]
\centering
\subfloat[Timestep 0.]{
\includegraphics[width=0.18\textwidth]{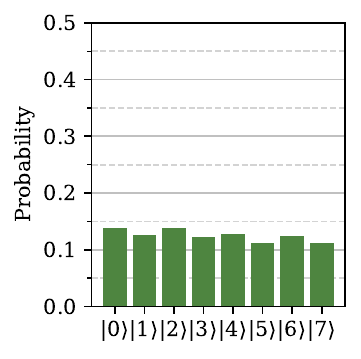}
\label{subfig_rwalks_q0}
}
\subfloat[Timestep 1.]{
\includegraphics[width=0.18\textwidth]{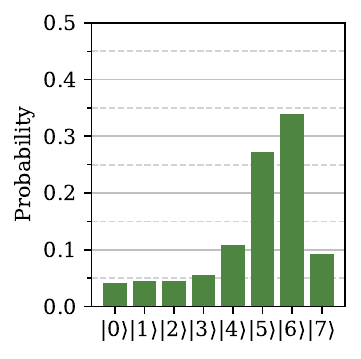}
\label{subfig_rwalks_q1}
}
\subfloat[Timestep 2.]{
\includegraphics[width=0.18\textwidth]{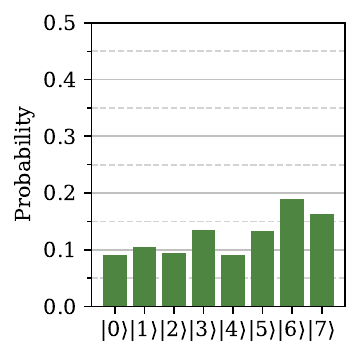}
\label{subfig_rwalks_q2}
}
\subfloat[Timestep 3.]{
\includegraphics[width=0.18\textwidth]{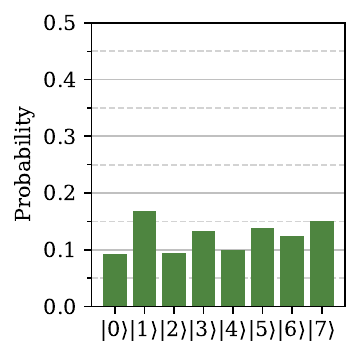}
\label{subfig_rwalks_q3}
}%
\subfloat[Timestep 4.]{
\includegraphics[width=0.18\textwidth]{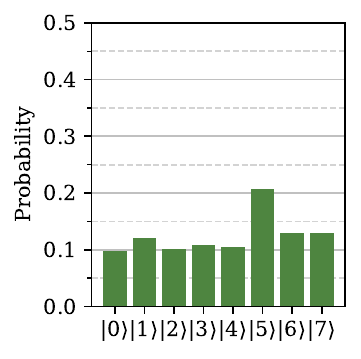}
\label{subfig_rwalks_q4}
}
\caption{Reflecting boundary \textit{Quantum Walk} on IBM Perth.}
\label{fig_refwalk_q}
\end{figure*}

\vspace{1mm}
\noindent
\textbf{Platforms: }We choose publicly available hardware platforms with mature software to evalaute \apps. 
For our gate-based applications, we use IBM systems (IBM Perth/Quito, which have 5 superconducting transmon qubits), and for annealing, we use D-Wave's Advantage\_system4.1 solver, which has 5,\,627 qubits connected in a Pegasus graph~\cite{dwaveSolver}.


\vspace{1mm}
\noindent
\textbf{Cognitive model parameters}: \apps can scale to the large sizes we target, but existing hardware cannot fit them. Therefore, we run smaller versions. 
Table~\ref{tab:impl_params} shows the minimum model sizes we'd like to evaluate, the actual sizes that we could fit on hardware, and the implementation parameters.

\begin{table}[ht]
\vspace{-2mm}
    \caption{Model and implementation parameters.} 
    \label{tab:impl_params}
    \renewcommand*{\arraystretch}{1}
    \scriptsize
    \centering
    \begin{tabulary}{\linewidth}{@{}p{0.1\linewidth}p{0.17\linewidth}p{0.17\linewidth}p{0.38\linewidth}@{}}
    \toprule
    & \multicolumn{2}{c}{\textbf{Model parameters}}  & \\
    \cmidrule(lr){2-3}
        \textbf{Model} & Ideal minimum & Actual & \textbf{Implementation parameters}\\ \toprule
        
        \multirow{2}{*}{\parbox{1cm}{Quantum Walk}} & \multirow{2}{*}{\parbox{1.72cm}{100 states, 1000 walks}} & 8 states, 1 walk & Gates. Qubits: 3 (ref), 4 (abs)\\
        \cmidrule(lr){3-4}
        && 4 states, 1 walk  & Annealing ($r$:2--9, 60\% pausing)\\ \midrule
        \multirow{2}{*}{MPMW} & \multirow{2}{*}{\parbox{1.72cm}{10 dynamic wells, 1000 positions, 10 eigenstates}} & \multirow{2}{*}{\parbox{1.72cm}{1 static well, 4 positions, 3 eigenstates}} & SSVQE (COBYLA/SPSA)  \\
        \cmidrule(lr){4-4}
        && & QITE (step size = 0.2, steps = 135) \\ \midrule
        Predator-Prey & 1000$\times$1000 grid, 100 levels, 10 icons & 6$\times$6 grid, 12 levels, 3 icons & RBM (112 node; data: 100 train, 100 test), 60\% pausing \\ \midrule
        LCA & 10 units, multilayer & 2 units, 1 layer & 5-step unroll, 6-bit floats, 60\% pausing \\
        \bottomrule
    \end{tabulary}
    \vspace{-1mm}
\end{table}


A large \textit{Quantum Walk} uses \textgreater100 states, and multiple walks to model interacting decisions. The smaller version we evaluate uses a single walk with 8 states for gate systems and 4 states for annealing. Annealing has fewer states since the actual number of qubits needed is much higher ($\ge$rN; Section~\ref{models}). We pause anneals~\cite{LCA_Pause} to mitigate thermalization and noisy outputs.

\textit{MPMW} scales to multiple wells and well-sizes, several eigenstates, and fine spatial resolution in the horizontal dimension. On actual hardware, we find 3 eigenstates for the simplest \textit{MPMW} model, which has one infinite-height well. In finding eigenvalues with SSVQE, we use the SPSA optimizer~\cite{qiskit} in simulations as it is more accurate, but use COBYLA~\cite{qiskit} on actual hardware since it requires fewer executions.


\textit{Predator-Prey} could scale to large screen grids, many attention levels, and icons. For the smaller version we run, our RBM has 56 visible and 56 hidden nodes.  The nodes take binary values, but their biases and the weights for the edges can have real values. Since we do not need the perceived positions explicitly, we use a single set of nodes for the true and perceived positions, encoded with one-hot vectors.

The RBM is trained with contrastive divergence~\cite{contrastiveDiv}. We obtain training data by identifying the best decision from an exhaustive search without lookahead. We train on 100 randomly chosen screens over 30 epochs, and use another 100 as test. While this data may seem small, 
evaluating it consumed hours of anneals and thousands of dollars.
 We use one-hot encoding to represent the screen grid coordinates, and allow the agent to move one step in any of the 8 cardinal and ordinal directions. 
 

A large \textit{LCA} model would have multiple layers, each with several LCA units to model cognitive control and decision-making. In our experiments, we use a single layer with 2 units. We use 6 qubits to represent float values. Since the LCA has multiple outputs over multiple timesteps, we quantify accuracy using the mean local relative error, which is the average error of a method $\overline{f}$ relative to exact LCA ($f$), when they run on the same inputs $i$ and previous values ($\overline{x_{-1}}$) from the method. It is given by, $avg(|\frac{f(i,{x_{-1}}) - \overline{f(i,{x_{-1}}})}{f(i,{x_{-1}})}|)$.


\section{Evaluation}
\label{results}

\subsection{Quantum Walk}
\label{sub:walkres}

\begin{figure*}[t]
\centering
\subfloat[Timestep 0.]{
\includegraphics[width=0.18\textwidth]{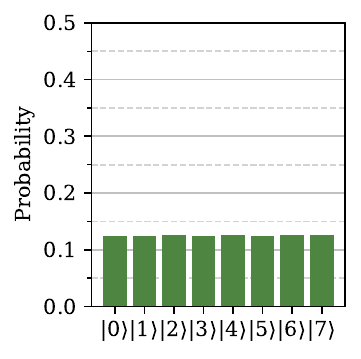}
\label{subfig_awalks_qasm0}
} 
\subfloat[Timestep 1.]{
\includegraphics[width=0.18\textwidth]{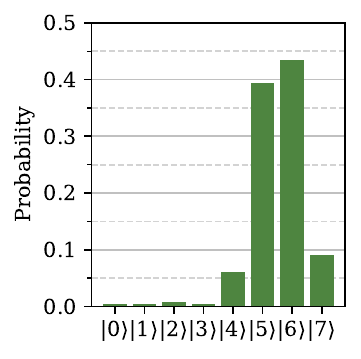}
\label{subfig_awalks_qasm1}
} 
\subfloat[Timestep 2.]{
\includegraphics[width=0.18\textwidth]{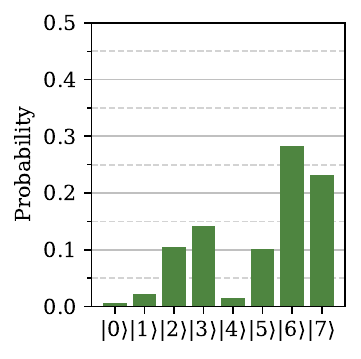}
\label{subfig_awalks_qasm2}
} 
\subfloat[Timestep 3.]{
\includegraphics[width=0.18\textwidth]{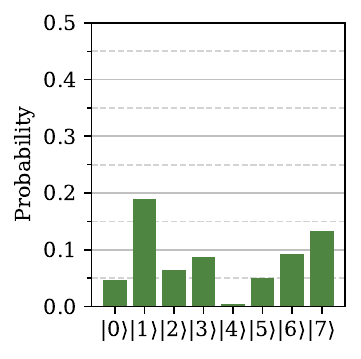}
\label{subfig_awalks_qasm3}
}
\subfloat[Timestep 4.]{
\includegraphics[width=0.18\textwidth]{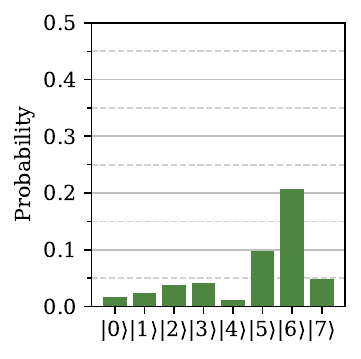}
\label{subfig_awalks_qasm4}
}
\caption{Absorbing boundary \textit{Quantum Walk} with statevector simulation (Total probability $<$1 due to projection).}
\label{fig_abswalk_qasm}
\vspace{-4mm}
\end{figure*}

\begin{figure*}[t]
\centering
\subfloat[Timestep 0.]{
\includegraphics[width=0.18\textwidth]{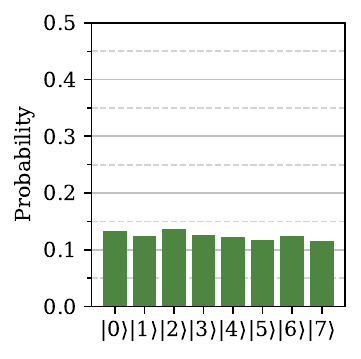}
\label{subfig_awalks_q0}
} 
\subfloat[Timestep 1.]{
\includegraphics[width=0.18\textwidth]{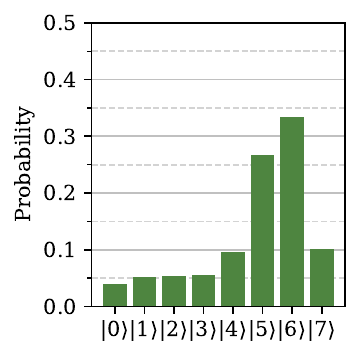}
\label{subfig_awalks_q1}
} 
\subfloat[Timestep 2.]{
\includegraphics[width=0.18\textwidth]{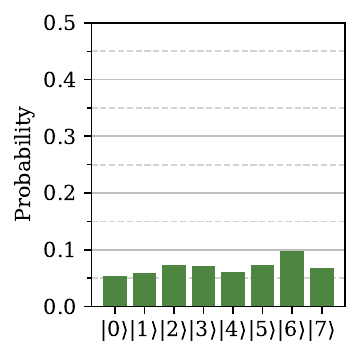}
\label{subfig_awalks_q2}
} 
\subfloat[Timestep 3.]{
\includegraphics[width=0.18\textwidth]{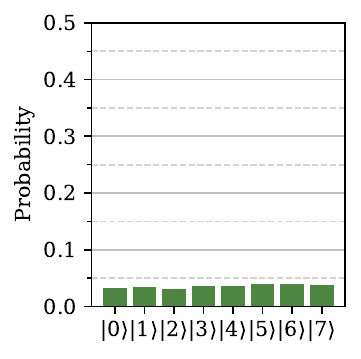}
\label{subfig_awalks_q3}
}
\subfloat[Timestep 4.]{
\includegraphics[width=0.18\textwidth]{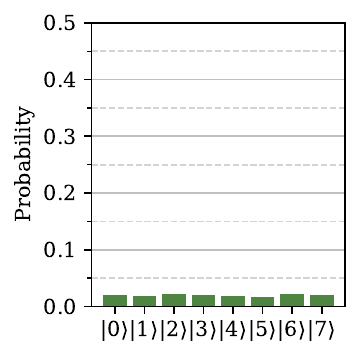}
\label{subfig_awalks_q4}
}
\caption{Absorbing boundary \textit{Quantum Walk} on IBM Perth (Total probability $<$1 due to projection).}
\label{fig_abswalk_q}
\end{figure*}

\noindent
\textbf{Gate-based implementation: }Figures~\ref{fig_refwalk_qasm} and~\ref{fig_refwalk_q} show the probabilities of the states at various timesteps in the reflecting boundary \textit{Quantum Walk} using statevector simulations, and as measured from quantum hardware (IBM Perth), respectively. The simulations align closely with our analytical calculations, but the results from quantum hardware deviate significantly, especially from timestep 2. 
The absorbing boundaries walk, shown in Figures~\ref{fig_abswalk_qasm} (simulations) and~\ref{fig_abswalk_q} (measured) is worse affected. 
Much of the distribution is lost even by timestep 2. 

\begin{figure}[h]
\vspace{-6mm}
\centering
\subfloat[Gate count.]{
\includegraphics[width=0.31\linewidth]{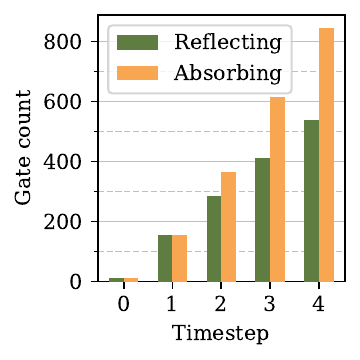}
\label{subfig_gatewalk_size}
}
\subfloat[Circuit depth.]{
\includegraphics[width=0.31\linewidth]{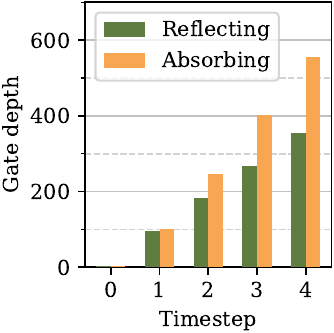}
\label{subfig_gatewalk_depth}
}
\subfloat[Execution time.]{
\includegraphics[width=0.31\linewidth]{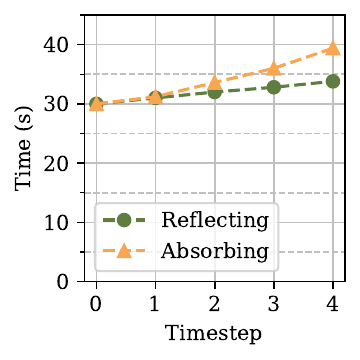}
\label{subfig_gatewalk_time}
}
\caption{Gate count, depth, and performance of \textit{Quantum Walk}.}
\label{fig_walkperf}
\vspace{-2mm}
\end{figure}
\begin{figure}[h]
\vspace{-6mm}
\centering
\subfloat[$p=0.01$]{
\includegraphics[width=0.31\linewidth]{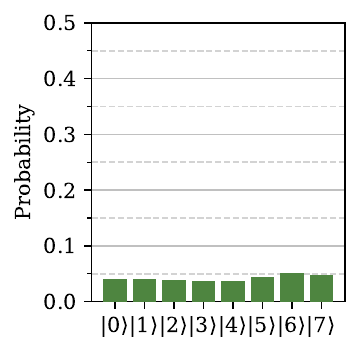}
\label{subfig_noise_p2}
}
\subfloat[$p=0.001$]{
\includegraphics[width=0.31\linewidth]{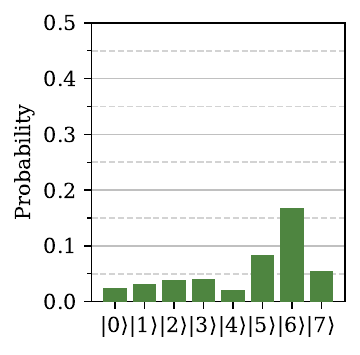}
\label{subfig_noise_p3}
}%
\subfloat[$p=0.0001$]{
\includegraphics[width=0.31\linewidth]{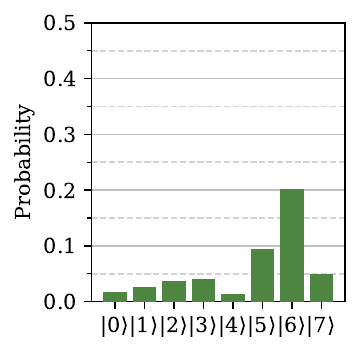}
\label{subfig_noise_p4}
}
\caption{Timestep 4 of absorbing walk with depolarizing noise.}
\label{fig_walknoise}
\vspace{-5mm}
\end{figure}

The cause of poor hardware performance is noise from running large circuits. 
Figure~\ref{fig_walkperf} shows the gate count, depth, and execution time of the walks. While these models use only 3 or 4 qubits, the gate depths are in the hundreds, resulting in large noise. We also simulate the absorbing boundary walk with 1- and 2-qubit gate depolarizing noise to study its sensitivity. Figure~\ref{fig_walknoise} shows the results at timestep 4 for different noise probabilities. There is noticeable distortion (vs Figure~\ref{subfig_awalks_qasm4}) at $p$=0.001, suggesting significant need for fault tolerance~\cite{campbell2017roads}, but near-accurate result for $p$=0.0001, indicating promise. 
\begin{tcolorbox}[colback=red!5!white,colframe=red!75!black,size=fbox,beforeafter skip=0pt]
Noise mitigation is vital for \textit{Quantum Walk} models, even more than building qubit capacity.
\end{tcolorbox}


\begin{figure}[h]
\vspace{-4mm}
\centering
\subfloat[Accuracy.]{
\includegraphics[width=0.31\linewidth]{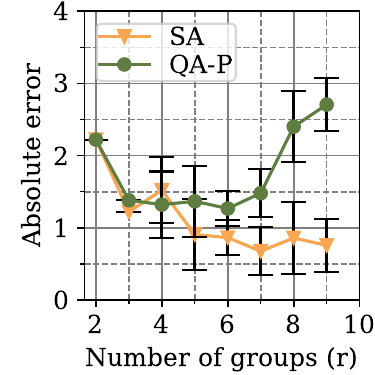}
\label{subfig_annealwalk_acc}
}
\subfloat[Size.]{
\includegraphics[width=0.32\linewidth]{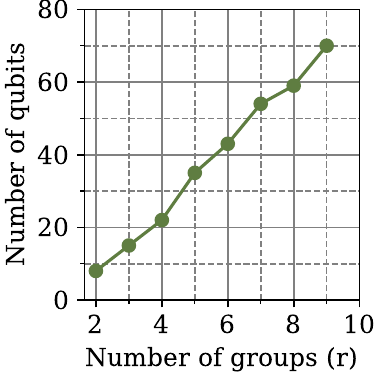}
\label{subfig_annealwalk_size}
}%
\subfloat[Execution time.]{
\includegraphics[width=0.31\linewidth]{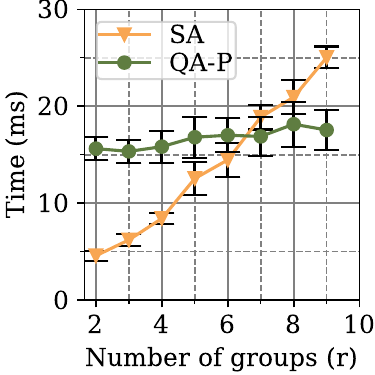}
\label{subfig_annealwalk_time}
}
\caption{Annealing for a 4-state Quantum Walk.}
\label{fig_annealwalkres}
\vspace{-3mm}
\end{figure}

\vspace{1mm}
\noindent
\textbf{Annealing implementation: } Figure~\ref{fig_annealwalkres} shows the results for finding the ground state of a 4-state \textit{Quantum Walk} using annealing. We show the results for both simulated annealing (SA) and quantum annealing with pausing (QA-P). Figure~\ref{subfig_annealwalk_acc} shows the absolute error of the measured eigenvalue (the exact value is -7.22). We expect the error to decrease as the qubit repetition ($r$) is increased, and we find this to be true for SA. With QA-P, however, the error increases after an initial decrease. This is because bigger $r$ values result in more qubits (Figure~\ref{subfig_annealwalk_size}) and complex interaction, worsening noise.

Figure~\ref{subfig_annealwalk_time} shows the execution time of SA and QA-P. SA's execution time grows linearly with $r$ while QA-P is relatively flat. Even though the error for QA-P is higher at large $r$, the trend for small $r$ (where its accuracy is comparable to SA) suggests that QA's execution time can scale better. 
\begin{tcolorbox}[colback=red!5!white,colframe=red!75!black,size=fbox,beforeafter skip=0pt]
QA suffers from noise with large \textit{Quantum Walk} models. With small models, QA's performance scales better over SA. \\ There's also an accuracy-time tradeoff to explore in annealers.
\end{tcolorbox}


\subsection{MPMW}
\label{sub:mpmwres}

\begin{figure}[h]
\vspace{-1mm}
\centering
\includegraphics[width=0.65\linewidth]{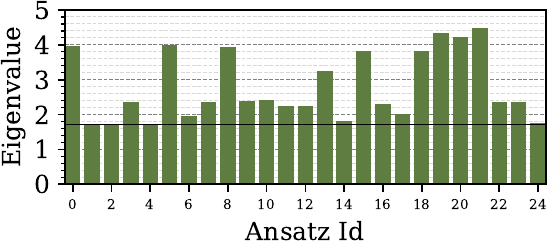}
\caption{Ansatz selection with VQE (the exact value is shown).}
\label{ssvqe_ansatz}
\end{figure}

\noindent
\textbf{SSVQE: } Identifying a suitable ansatz for SSVQE was nontrivial. We simulated 24 choices from prior work~\cite{ansatzChoice,variationalAlgo,kandala2017hardware} to solve for \textit{MPMW}'s ground state using standard VQE, and 
Figure~\ref{ssvqe_ansatz} shows their performance. 
The best choice was (Ansatz 1), which is ``Circuit 14'' from~\cite{ansatzChoice} that uses $R_Y$ and controlled $R_X$ gates. We also used circuit search methods from quantum ML~\cite{quantumnas}, but they did not perform competitively.  

 For optimizers, we evaluated both SPSA and COBYLA in our simulations, and found SPSA to be better. However, SPSA required 2--10$\times$ more circuit evaluations,  which we could not run on real hardware in reasonable time for convergence. So, we only used COBYLA for quantum hardware. 

\begin{figure}[h]
\vspace{-5mm}
\centering
\subfloat[Simulated.]{
\includegraphics[width=0.37\linewidth]{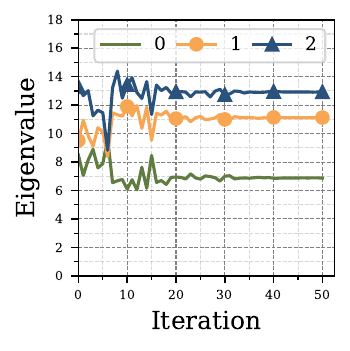}
\label{subfig_ssvqe_cobylasim_iter}
} 
\subfloat[Measured.]{
\includegraphics[width=0.37\linewidth]{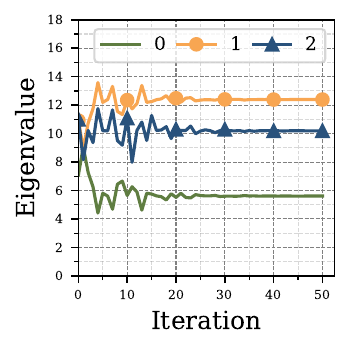}
\label{subfig_ssvqe_cobylaq_iter}
} 
\caption{Simulated and measured SSVQE B2 for \textit{MPMW}.}
\label{fig_ssvqeiters}
\vspace{-3mm}
\end{figure}

We run the SSVQE B and SSVQE C algorithms to find the 0$^\textrm{th}$, 1$^\textrm{st}$ and 2$^\textrm{nd}$ eigenstates of \textit{MPMW}, with exact values as 1.72, 6.21, and 11.78, respectively. Figure~\ref{fig_ssvqeiters} shows the progression of SSVQE B2, which only finds the 2$^{nd}$ eigenstate. The measured value (10.1) has a steady state error (14\%). Other runs (SSVQE B0/B1/C) are similar. 

\begin{figure}[h]
\vspace{-5mm}
\centering
\subfloat[SPSA.]{
\includegraphics[width=0.31\linewidth]{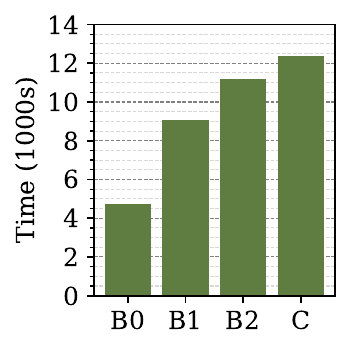}
\label{subfig_ssvqespsa_time}
}
\subfloat[COBYLA.]{
\includegraphics[width=0.31\linewidth]{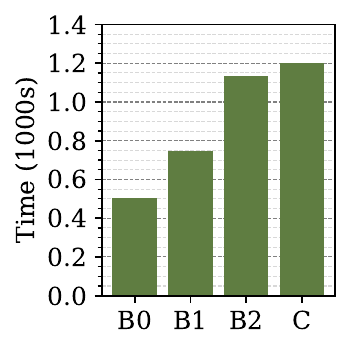}
\label{subfig_ssvqecobyla_time}
}
\subfloat[COBYLA (IBM Quito).]{
\includegraphics[width=0.31\linewidth]{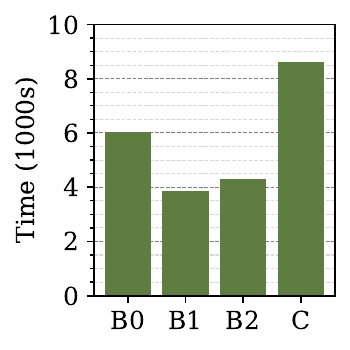}
\label{subfig_ssvqecobyla_q_time}
} 
\caption{Execution time of the SSVQE algorithm for \textit{MPMW}}
\label{fig_ssvqetimes}
\vspace{-4mm}
\end{figure}

Figure~\ref{fig_ssvqetimes} shows the execution time of SSVQE. SSVQE uses many circuit evaluations to compute the Pauli expectations, repeated for each eigenstate, causing long execution times. The evaluations increase  with the SPSA optimizer. 


Figure~\ref{fig_ssvqetimes} also shows that each SSVQE B instance takes lesser time than SSVQE C, which finds all eigenvalues at once. Thus, if these multiple instances could be run in parallel, the time to solution will be faster. Unfortunately, existing quantum clouds do not support such execution.
\begin{tcolorbox}[colback=red!5!white,colframe=red!75!black,size=fbox,beforeafter skip=0pt]
Little support for ansatz and optimizer selection.\\
There is embarassing parallelism to be exploited in SSVQE.
\end{tcolorbox}

\begin{figure}[h]
\vspace{-3mm}
\centering
\subfloat[Convergence.]{
\includegraphics[height=0.4\linewidth]{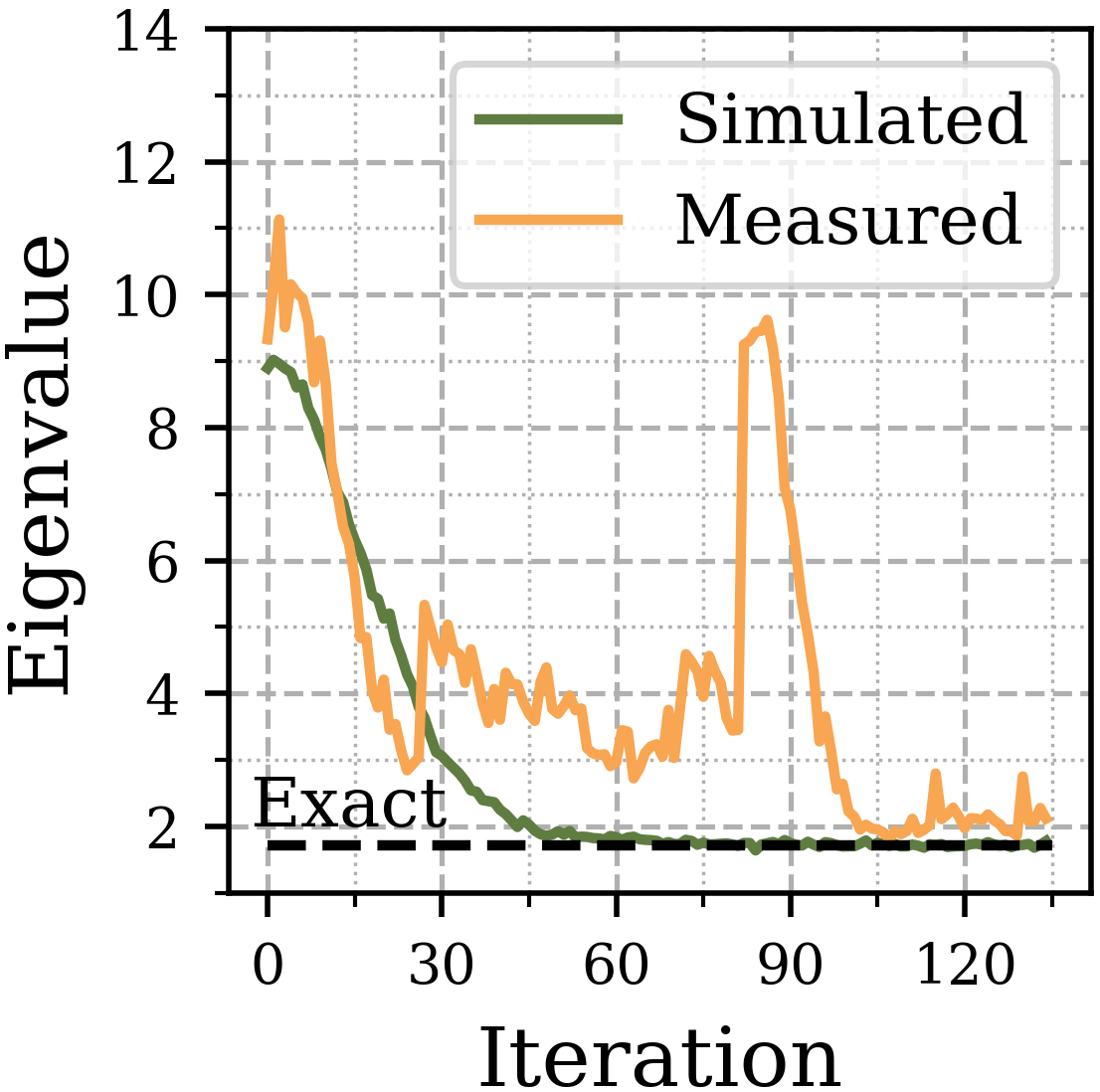}
\label{subfig_qite_converge}
} 
\subfloat[Accuracy of eigenstates.]{
\includegraphics[height=0.4\linewidth]{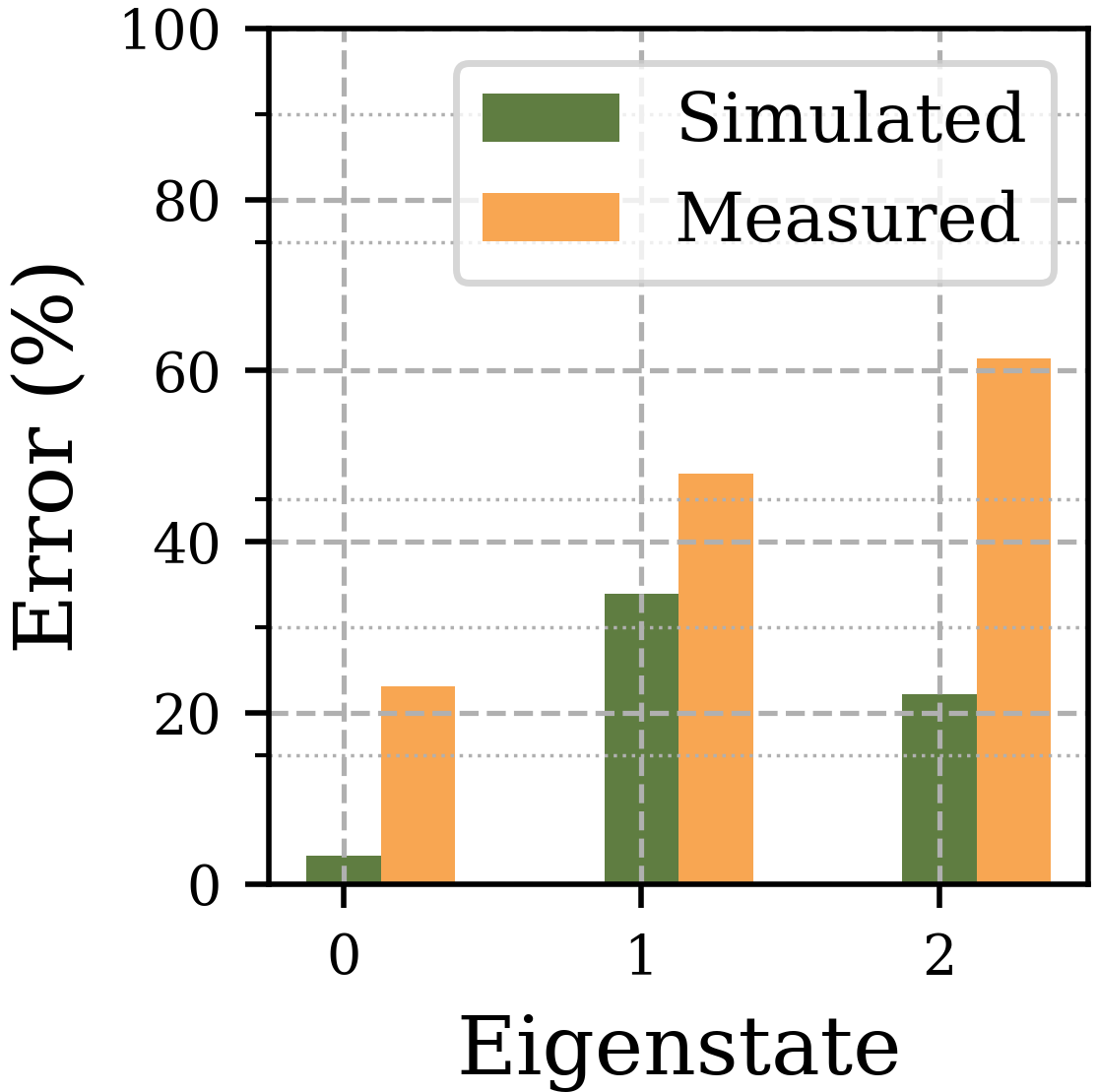}
\label{subfig_qite_acc}
} 
\caption{QITE and Quantum Lanczos for \textit{MPMW}.}
\label{fig_qiteres}
\vspace{-2mm}
\end{figure}

\vspace{2mm}
\noindent
\textbf{QITE/Quantum Lanczos: } Figure~\ref{fig_qiteres} shows their performance for \textit{MPMW}. 
Figure~\ref{subfig_qite_converge} shows convergence of the ground state but also significant hardware noise. Figure~\ref{subfig_qite_acc} shows the error in the calculated eigenvalues. The ground state error is low and is better than what SSVQE achieved. The excited states have a steady state error due to our unitary approximation, but the error is only slightly worse than with SSVQE. 
Importantly, however, QITE is much faster than SSVQE, despite being nonunitary. It required just 164.8\,s in simulations (vs 12,000\,s for SSVQE C with SPSA for comparable accuracy). This is because it has fewer evaluations, and doesn't tune an ansatz. Moreover, QITE/Quantum Lanczos do not require an ansatz and optimizer search, which is hard. Unfortunately, these methods haven't received adequate attention in systems.
\begin{tcolorbox}[colback=red!5!white,colframe=red!75!black,size=fbox,beforeafter skip=0pt]
Non-unitary methods are competitive or better than variational ones for near-term exploration, but lack systems support.
\end{tcolorbox}

\subsection{Predator-Prey}
\label{sub:ppres}

\begin{figure}[h]
\vspace{-4mm}
\centering
\subfloat[Multiple anneals (RBM).]{
\includegraphics[width=0.31\linewidth]{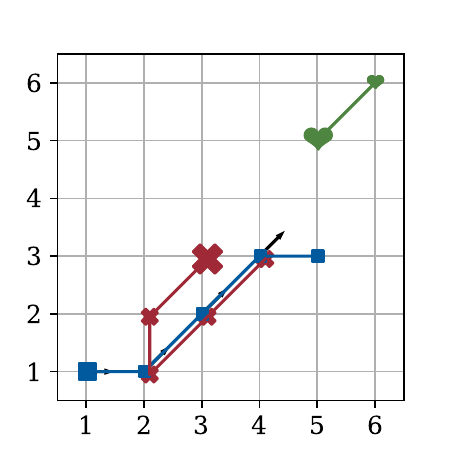}
\label{subfig_pp_qa}
} 
\subfloat[Sample-efficient anneal (RBM$_\textrm{eff}$).]{
\includegraphics[width=0.31\linewidth]{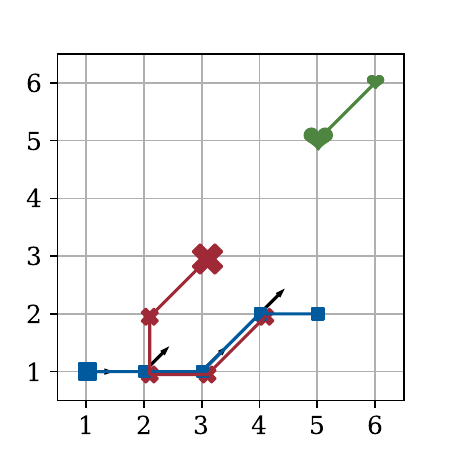}
\label{subfig_pp_combined}
} 
\subfloat[Two-sample parallel anneal (RBM$_\textrm{par2}$).]{
\includegraphics[width=0.31\linewidth]{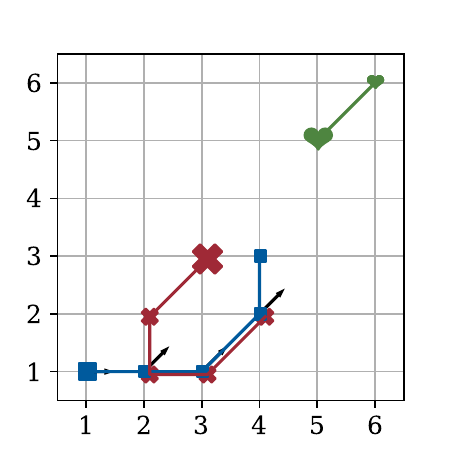}
\label{subfig_pp_parallel}
} 
\caption{\textit{Predator-Prey} screen with different RBM  strategies.}
\label{fig_ppres}
\vspace{-4mm}
\end{figure}

Figure~\ref{fig_ppres} shows a few sequences of moves generated by our RBM with the three different annealing strategies (RBM, RBM$_\textrm{eff}$, and RBM$_\textrm{par2}$), and the corresponding movements from the predator and prey. At each step, we show the deviation of the RBM's move from a no-lookahead exhaustive search. 

\begin{figure}[h]
\vspace{-4mm}
\centering
\subfloat[Timing.]{
\includegraphics[width=0.4\linewidth]{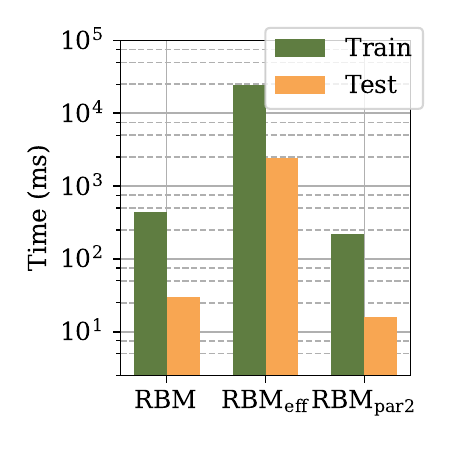}
\label{subfig_pp_timing}
} 
\subfloat[Accuracy.]{
\includegraphics[width=0.4\linewidth]{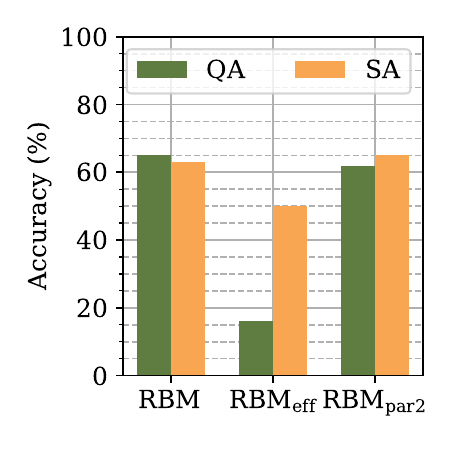}
\label{subfig_pp_acc}
} 
\caption{\textit{Predator-Prey} performance.}
\label{fig_pp_perf}
\vspace{-3mm}
\end{figure}

Figure~\ref{subfig_pp_timing} shows the execution time of our RBM implementations.  Compared to RBM, RBM$_\textrm{eff}$ takes nearly 100$\times$ longer. This is because RBM$_\textrm{eff}$  packs 2 passes into one anneal resulting in significantly complex interactions, and thus, noise. The number of physical qubits to realize this network is 100$\times$ higher. On the other hand, RBM$_\textrm{par2}$ is $\approx$2$\times$ faster. Even though it packs the processing of 2 samples into one anneal, the two samples are processed independently and do not complicate qubit interactions.  Thus, it only needs 2$\times$ more qubits. 

Figure~\ref{subfig_pp_acc} shows the accuracy of the RBMs with simulated (SA) and quantum annealing (QA). RBM$_\textrm{eff}$ has poor accuracy due to its complex interactions, which is more pronounced with noise-sensitive QA. On the other hand, RBM$_\textrm{par2}$ has comparable accuracy to RBM. Note that in training, RBM$_\textrm{par2}$ updates the weights after 2 data samples, deviating from the standard algorithm that updates weights after each sample. However, this doesn't affect the RBM's accuracy, similar to what has been observed in some classical ML training algorithms~\cite{hogwild}. We, thus, find a new opportunity to accelerate quantum RBMs. But, existing quantum clouds don't support such parallel execution, and we had to emulate it with 0-strength qubit couplings.
\begin{tcolorbox}[colback=red!5!white,colframe=red!75!black,size=fbox,beforeafter skip=0pt]
There is data-parallelism to be exploited with quantum RBMs.
\end{tcolorbox}


\subsection{LCA}
\label{sub:lcares}

\begin{figure}[h]
\vspace{-2mm}
\centering
\includegraphics[width=0.9\linewidth]{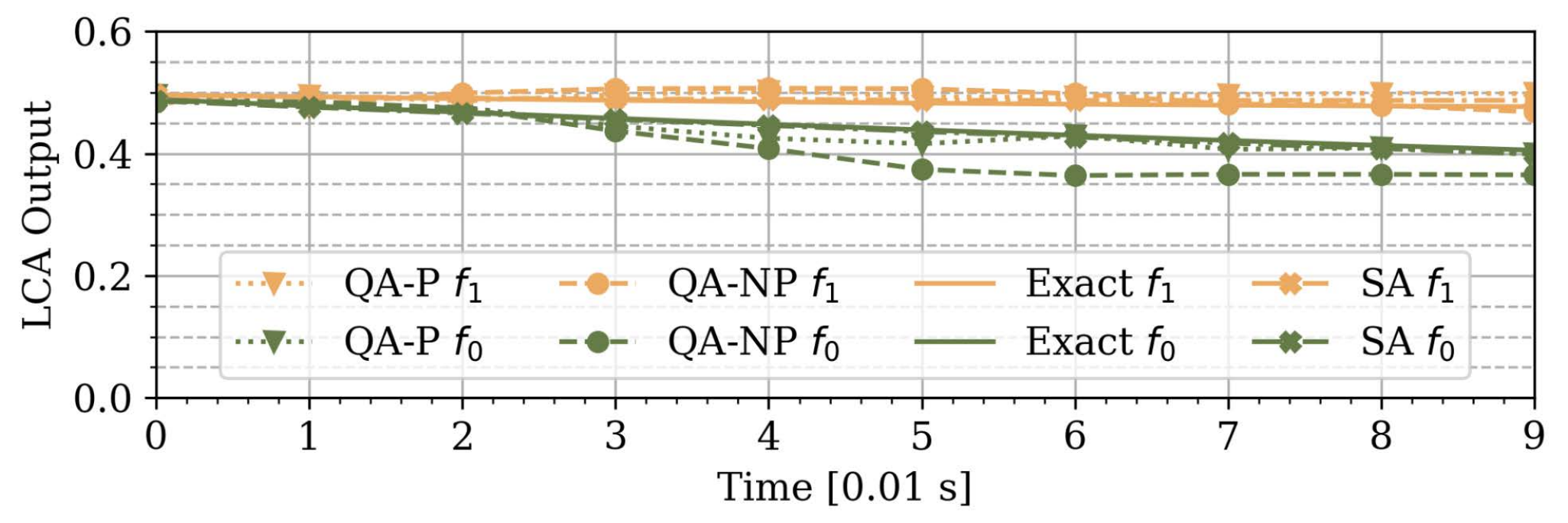}
\caption{Quantum annealing for the LCA.}
\label{fig_lcares}
\vspace{-3mm}
\end{figure}

Figure~\ref{fig_lcares} shows the two outputs of LCA ($f_1$, and $f_2$) for 10 timesteps, computed with different methods: SA, QA-P, and QA-NP (QA without pausing). We find 5 timesteps in one anneal, requiring 60 logical qubits and 252 physical qubits. 

\textit{LCA} with SA is close to the exact values, with a mean local error (Section~\ref{setup}) of (0.58\%, 0.49\%) for the two LCA outputs. Among the QA methods, QA-NP has larger deviations due to noise, with an error of (5.75\%, 7.54\%). QA-P has a better fit with an error of (5.18\%, 5.82\%).  Note that our error metric does not consider the conditioning of the problem i.e., for ill-conditioned calculations, errors are amplified over time.

For execution time, SA took 84.26\,ms for one anneal with 100 reads and 1000 sweeps. For the best fidelity, QA required 175\,ms  QPU (quantum processing unit) time for 3500 reads with 50\,$\mu$s anneals each. With simple noise mitigation (i.e., pausing) QA for the \textit{LCA} model achieved comparable accuracy to SA. However, QA also needed 3.5$\times$ more reads, and consequently, 2$\times$ longer time. Noise mitigation and device reliability could bridge this gap in the near term. 
\begin{tcolorbox}[colback=red!5!white,colframe=red!75!black,size=fbox,beforeafter skip=0pt]
Exiting annealers are noisy to run \textit{LCA}, but mitigation can offer superior performance in the near-term.
\end{tcolorbox}
\section{Case Study: Cloud Co-design}
\label{cloud}

We demonstrate research that \apps inspires through the design of efficient quantum clouds that support parallelism.

\begin{figure}[ht]
\vspace{-5mm}
\centering
\subfloat[Existing.]{
\includegraphics[width=0.44\linewidth]{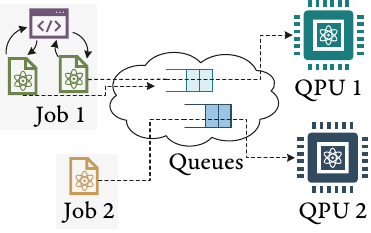}
\label{subfig_cloud_cur}
}
\subfloat[Proposed.]{
\includegraphics[width=0.5\linewidth]{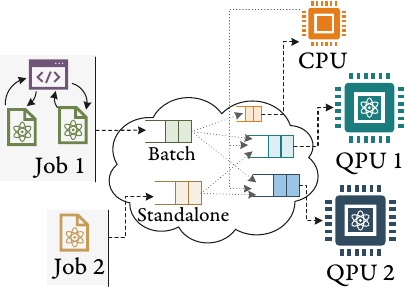}
\label{subfig_cloud_new}
} 
\caption{Reorganizing quantum cloud scheduling.}
\label{fig_cloudorg}
\end{figure}

Existing quantum clouds do not leverage the parallelism found in \apps, e.g., in eigensolvers, or RBMs. Furthermore, we persistently encountered long scheduling delays in our evaluation. These occur due to an inefficient cloud. Figure~\ref{subfig_cloud_cur} shows the scheduling on quantum clouds today~\cite{cloudSched1,dwaveQuantum}. Users place each job into device-specific queues, where they stay without migration until execution. This inflexibility combined with the ad hoc job execution heuristics~\cite{cloudSessions} results in load imbalance across devices, and long wait times for users. Iterative jobs like SSVQE or QITE additionally suffer a long network round trip between quantum and classical devices for every iteration.
The present design also fails to benefit from parallelism because when jobs like SSVQE spawn multiple parallel instances, they are all run serially on the same device. 

We envision a new cloud that (i) separates user and device scheduling, (ii) runs hybrid, and not just quantum jobs, and (iii) allows autonomous hybrid job execution with cloudlets~\cite{cloudlet}. Figure~\ref{subfig_cloud_new} shows our design. We distinguish user-facing front-end scheduling from back-end device-specific job selection. Users submit jobs to the front-end without selecting a specific device. Instead, they are mapped to any available device (considering preferences and priorities), reducing imbalance and wait times.  Next, to reduce long-latency network round trips, we propose adding cloudlets~\cite{cloudlet} with classical compute (CPUs/GPUs). Cloudlets are small, \textit{geographically close} servers, that have gained prominence in mobile network systems. Users submit both quantum and hybrid tasks onto this system, where the hybrid jobs run autonomously on respective devices. 

Existing systems lack the features (i)-(iii) above.
Amazon~\cite{awsQuantum} allows hybrid job scheduling, but users must select a single device. The hybrid jobs cannot also create additional instances for other devices, which is needed for dynamic parallelism. 



Our envisioned cloud is more aligned with the needs of the \apps applications. It enables exploiting the embarrassing parallelism that we find in eigensolvers and RBMs, \textit{without any hardware change or synchronization}, complementing other forms of parallelism explored in prior work~\cite{simultaneousPauli,simultaneousPauli2,mineh2023accelerating,stein2022eqc}.

\begin{figure}[h]
\vspace{-4mm}
\centering
\subfloat[\textit{MPMW} SSVQE.]{
\includegraphics[width=0.3\linewidth]{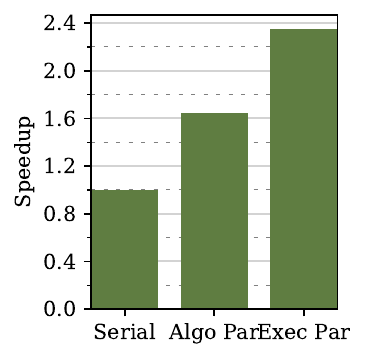}
\label{subfig_ssvqe_cloud}
} 
\subfloat[Parallel Predator-Prey (time).]{
\includegraphics[width=0.3\linewidth]{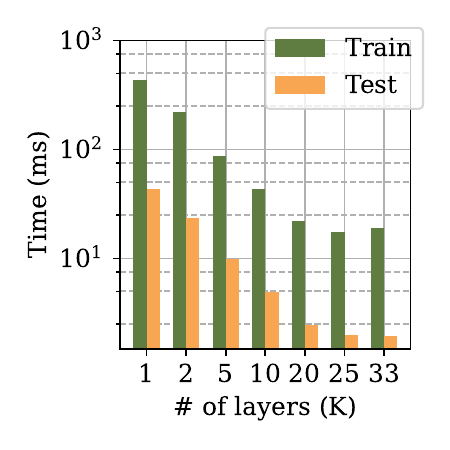}
\label{subfig_rbm_cloudtime}
} 
\subfloat[Parallel Predator-Prey (accuracy).]{
\includegraphics[width=0.3\linewidth]{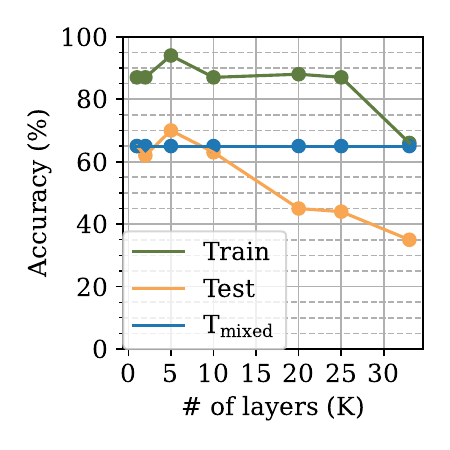}
\label{subfig_rbm_cloud}
} 
\caption{Impact of a parallelism-aware quantum cloud.}
\label{fig_cloud}
\vspace{-3mm}
\end{figure}

Figure~\ref{subfig_ssvqe_cloud} shows the normalized speedup for \textit{MPMW} that can be obtained by executing the 3 instances of SSVQE B in parallel, or one SSVQE C instance  (algorithmic multitasking/parallelism) vs serially running SSVQE B (current option), even without cloudlets. Parallel execution yields $\approx$2.4$\times$ speedup with no design effort.  When other forms of parallelism are combined, the improvements would be multiplicative. 

For the RBM in the \textit{Predator-Prey} model, Figure~\ref{subfig_rbm_cloudtime} shows the training and testing times, and Figure~\ref{subfig_rbm_cloud} shows the accuracy when processing varying number of samples in parallel on a system that permits such execution. 
We also show a design that is trained in serial, but tested in parallel (T$_\textrm{mixed}$). The RBM's accuracy remains relatively unchanged up to 10-parallel evaluations during training, but is 10$\times$ faster than serial execution. Furthermore, testing can always run in parallel without loss of accuracy. These results show the significant benefits that a new cloud inspired from \apps  can provide with no change in hardware or toolchains.  


\section{Lessons Learnt}
\label{analysis}

The design and execution of \apps applications required innovation, and has also revealed new research. Table~\ref{tab:takeaways} organizes these findings. In addition to the discussion on organization in section~\ref{cloud}, we briefly discuss the rest here. 

\begin{table}[ht]
\vspace{-2mm}
    \caption{Innovation and limitations exposed.} 
    \label{tab:takeaways}
    \renewcommand*{\arraystretch}{1}
    \scriptsize
    \centering
    \begin{tabulary}{\linewidth}{@{}p{0.13\linewidth}p{0.19\linewidth}p{0.23\linewidth}p{0.14\linewidth}p{0.12\linewidth}@{}}
    \toprule
         & \textbf{Algorithms} &\textbf{ Compilers \& Tools} & \textbf{Organization} & \textbf{Hardware}\\ \toprule
\textbf{Exposed limitations} & Few algorithms for analog quantum computers, nonlinear dynamics & Limited support for projections, analog quantum computers, non-unitary algorithms, ansatz search & Poor scheduling, lack of support for parallelism & Limited capability, noise \\ \midrule

 \textbf{Innovation \& discovery} & State-detector, Data-parallel RBMs, parallelism in eigensolvers & Hamiltonian walks on annealers & Cloud co-design for parallelism & \\
        \bottomrule
    \end{tabulary}
    \vspace{-1mm}
\end{table}

\noindent
\textbf{Software (Algorithms, Compilers): } Present quantum software frameworks are heavily specialized (e.g.,~\cite{qforte,openfermion,Kottmann2021Mar,Luo2020Oct}). We explored them for \apps but the abstractions mismatch. 
The toolchains  are also heavily focused on specific computation-styles (unitary or Ising Hamiltonians), devices (gate-based or annealers) and algorithms (variational). There is little tool support for competitive alternatives like Hamiltonian-based quantum systems~\cite{simuq,queraQuantum}, or non-unitary algorithms. Even popular constructs like ansatz search, or partial projectors lack generalizable tools. This reduces programmability and restricts the applications that can be run on quantum systems today. 

We also identify new research into the asynchronous training of RBMs. Understanding the theoretical mechanisms of this approach, and its applicability for other quantum ML networks would be valuable, as it did in classical computing~\cite{hogwild}.

Lastly, the new cloud design we envision also calls for new runtime managers and quantum operating systems to provide performance and protection.




\vspace{1mm}
\noindent
\textbf{Hardware-Software interfaces: }Existing hardware exposes device control parameters such as the pulse information required to implement gates~\cite{qiskitPulse}, or the annealing schedules for annealers.  While this enables cross-layer co-design~\cite{gokhale2020optimized}, it also creates additional programming burden to identify the correct settings---e.g., anneal schedules for pausing in the \textit{LCA}. This is especially hard with the limited tools available today.

Striking a balance, we propose that device control mechanisms be abstracted with compiler heuristics. For example, pausing can be automated by using recent work that can generate a quick estimate~\cite{imoto2022guaranteed}.


\vspace{1mm}
\noindent
\textbf{Hardware: } While current NISQ (noisy intermediate-scale quantum) devices may not outperform the best classical systems, we show promising empirical results that the \apps models could do so in the near future. We highlight (in Section~\ref{sub:walkres}) that, with improved noise rate (at $10^{-4}$), we can already start to explore quantum walk models accurately, representing meaningful NISQ or early fault tolerant applications. 

We also argue for research increasing device \textit{functionality}, complementing the present focus on noise resilience and qubit counts. 
Lack of functionality, e.g., the absence of primitives to express projections in \textit{Quantum Walk}, requires users to design custom circuits and pre-/post-selection methods. This is tedious, error-prone since quantum programming is not intuitive~\cite{ppsParadox}, and inefficient since it wastes qubits and execution time. While capabilities towards these features exist~\cite{monroe2021partial}, hardware development inevitably takes long, and such features could at least be software emulated in the interim. 

\section{Related Work}
\label{relwork}

\noindent
\textbf{Quantum applications: } Numerous advances~\cite{Arute2019Oct,Zhong2020Dec,Wu2021Oct,Madsen2022Jun,Bharti2022Feb,Zhang2023Jul} have helped make quantum computing a reality. However, even though there is great interest in adopting quantum computing for new domains (e.g.,~\cite{ionqOrder,quantBio}), realizing new applications  has been a challenge~\cite{academiesQuantum,quantumDiscovery}. Applications are focused in only a few areas~\cite{feynman1982simulating, Aspuru-Guzik2005Sep,quantumFinance,quantumML,quantumOpt}. We identify cognitive modeling as a new domain that can benefit from quantum computing, and present a suite of real-world applications. These are full models unlike benchmark suites, which use kernels for performance comparisons~\cite{supermarq,lubinski2023application,finvzgar2022quark,ionqOrder}.  Our work also spans analog and digital quantum computers, which is rare.  

\noindent
\textbf{Quantum computer architecture: } Significant strides have been made in quantum programming and compilation~\cite{scaffcc}, circuit synthesis~\cite{synQuant,fastOptSyn}, noise mitigation and reliability~\cite{veritas,qraft,mitigateBias,crosstalk,afs,q3de}, and microarchitecture design~\cite{murali19insight}.  Recent work~\cite{das2021jigsaw} on exploring improving fidelity of partial measurements, and improving variational algorithms for near-term machines~\cite{cafqa,qismet} can help the models we study.  

One limitation of prior research is that it is heavily device-centric~\cite{veritas,qraft,mitigateBias,crosstalk,afs,q3de, scaffcc,cutqc,murali19insight}, with only a few analyses at the system level~\cite{quantumJobMan,cloudAnalysis,multiTask}. Our work on higher level organization complements existing research. 
Furthermore, existing systems research and software frameworks are almost entirely focused on gate-based systems, which are only one type of quantum computers. Many applications, like some of ours, are more naturally suited to Hamiltonian-based computers (e.g., QuEra~\cite{queraQuantum,dwaveQuantum}). Unfortunately, the stack for these types of systems is limited. 

Ravi et al.~\cite{quantumJobMan,cloudAnalysis} raised the issue of growing application demand in the cloud, and proposed fidelity- and queuing-aware scheduling. We consider other aspects like parallelism.

\noindent
\textbf{Parallelism and Scheduling: } Prior work explored microarchitecture and circuit-level parallelism with co-location on gate-based computers~\cite{microarchParallel,niu2022parallel,ohkura2022simultaneous,multiProgramGate,liu2021qucloud} and annealers~\cite{multiTaskAnneal,multiTask} while mitigating reliability issues that arise.
We study a complementary form of embarassing parallelism exposed by \apps, that can benefit from, but does not need co-location. Capturing all these forms of parallelism could enable near-term machines to be competitive for real applications. 

\section{Conclusion}

Quantum computing can benefit many domains, but it has been a challenge to identify new applications, and stimulate architecture-application co-design. This work presented cognitive modeling as a new application area for quantum computing. We developed \apps, a suite of  real-world cognitive models that can be run on existing quantum  hardware. Developing \apps required innovation, and running them helped us identify new research in the quantum stack, some of which we evaluate with real data. Our work simultaneously advances the cognitive sciences, and quantum computer architecture.

\balance

\bibliographystyle{IEEEtran}
\bibliography{references}

\end{document}